\pdfoutput=1
\documentclass{PoS}

\usepackage{graphicx}

\def\lsim{\raise0.3ex\hbox{$<$\kern-0.75em\raise-1.1ex\hbox{$\sim$}}}
\def\gsim{\raise0.3ex\hbox{$>$\kern-0.75em\raise-1.1ex\hbox{$\sim$}}}

\title{Finite Temperature QCD on the Lattice -- Status 2010}

\ShortTitle{Finite Temperature QCD on the Lattice -- Status 2010}

\author{\speaker{Kazuyuki Kanaya}\\
        Graduate School of Pure and Applied Sciences, University of Tsukuba,\\ 
        Tsukuba, Ibaraki 305-8571, Japan\\
        E-mail: \email{kanaya@ccs.tsukuba.ac.jp}}

\abstract{
In the last couple of years, there has been big progress in finite temperature QCD on the lattice.
Large-scale dynamical simulations of 2+1 flavor QCD with various improved staggered quark actions have been started to produce results for various thermodynamic quantities which are extrapolated to the continuum limit at around physical quark masses, 
and thus are capable for a direct comparison with experiment.
At the same time, the theoretical uneasiness with staggered-type lattice quarks motivated several groups to accelerate studies with Wilson-type quarks and lattice chiral quarks. 
In this review, I discuss these important developments in finite temperature QCD made in the past year.
}

\FullConference{The XXVIII International Symposium on Lattice Field Theory, Lattice2010\\
		June 14-19, 2010\\
		Villasimius, Italy}

\begin{document}

\section{Introduction}

The transition from the conventional hadronic matter to the quark-gluon plasma (QGP) is one of the most drastic predictions of QCD. 
The rapid progress of heavy ion collision experiments urge us towards  
quantitatively reliable determination of properties of hot quark matter
directly from the first principles of QCD.
In this note, I review the recent status of studies on the nature of the quark matter at finite temperatures on the lattice.

Study of finite temperature QCD often requires simulations on coarse lattices. 
It is thus important to keep in mind possible caveats from lattice artifacts. 
The choice of the lattice quark action has significant implications for the nature of lattice artifacts.
Therefore, before discussing individual topics, I first comment in Sect.~\ref{sec:LatticeQuarks} on the nature of various lattice quarks adopted in the studies of finite temperature QCD, as well as possible caveats from them.

Because the $s$ quark mass is comparable with the QCD scale $\Lambda_{\rm QCD}$ and thus the transition temperature $T_c$,
it is important to incorporate the dynamical effects of the $s$ quark for realistic predictions about QGP. 
In Sect.~\ref{sec:PhaseStructure}, I discuss the status of our understanding of the phase structure of QCD with 2+1 flavors of quarks. 
There was a conflict about the value of $T_c$ with improved staggered quarks. 
This year, the discrepancy is mostly resolved through intensive simulations on fine and large lattices.
From these developments, we learn the importance of a good control of lattice artifacts in order to achieve a consistent prediction for experiment.
Scaling is a powerful tool to discriminate the nature of the transition.
I discuss recent progress in the studies on the scaling properties of the QCD transition.

Section~\ref{sec:EOS} is devoted to the status of the equation of state (EOS) in 2+1 flavor QCD.
The scope of finite temperature QCD is quite wide. 
Some of important developments in various directions are discussed in Sect.~\ref{sec:HotIssues}.
Studies of finite density QCD will be summarized by S.\ Gupta in these proceedings \cite{SGuptaLat10}.
For recent reviews, see \cite{LaineLat09,DeTarLat08,FodorKarschLat07}.

\section{Quarks on the lattice}
\label{sec:LatticeQuarks}

\subsection{Staggered-type quarks}
\label{sec:KS}

Most quantitative outputs of finite temperature QCD have been produced with staggered-type lattice quarks.
We have good reasons to adopt these quarks.
Namely
(i) they preserve a kind of chiral symmetry (U(1) symmetry under a taste-chiral transformation) on the lattice, which protects the location of the chiral limit against quantum fluctuations, 
and
(ii) they are relatively cheap to simulate. 
The latter is important for studies of finite temperature QCD which usually require a systematic survey in a wide range of the parameter space.

On the other hand, the staggered quarks lead to four copies (``tastes'') of fermions for each flavor in the continuum limit.
To remove unwanted three tastes, we usually adopt the ``fourth root procedure'', i.e.\ we replace the quark determinant by its fourth root in the path-integration procedures.
Since this makes the theory non-local and non-unitary, many theoretical issues emerge \cite{rStag1,rStag2,rStag3,rStag4}.
In particular, the universality arguments become fragile due to non-local interactions, and thus, e.g.\ we are not fully sure if the staggered quarks have the continuum limit with desired number of flavors.
So far, no apparent obstacles about the validity of rooted staggered quarks seem to exist, provided that the continuum extrapolation is done prior to the chiral extrapolation \cite{rStag1,rStag4}.
It is out of the scope of this review to go into these discussions.
Therefore, in the followings, I simply assume that the continuum extrapolation is (will be) done first and that the staggered quarks do have the continuum limit which is in the desired universality class of the QCD with arbitrary number of flavors.

Even with these assumptions, a couple of issues remain. 
One of them is the problem of taste violation.
The symmetry among tastes in the continuum limit is explicitly violated at finite lattice spacings.
This introduces a systematic uncertainty in the identification of flavors.
Due to the taste violation, for example, we observe different masses for the same hadron depending on the combination of tastes for the hadron operator.
The largest taste violation is known to appear in the pions \cite{Ishizuka94}.
With the U(1) taste-chiral symmetry preserved on the lattice, one combination of tastes, the pseudo Nambu-Goldstone (pNG) pion, leads to the vanishing mass in the chiral limit, but other 15 combinations remain heavy.
Although the lightest pNG pion is customary treated as the pion, heavier pions do contribute to loop diagrams and thus induce lattice artifacts.
To reduce the magnitude of lattice artifacts, various improved actions have been proposed, such as asqtad \cite{asqtad}, HYP \cite{HYP}, p4 \cite{p4}, stout \cite{stout}, HISQ \cite{HISQ}, etc.
Each of them may be originally designed to improve different quantities, such as the free quark dispersion relation, perturbative propagators  etc., but reduces the taste violation too directly or indirectly.
Note that, unlike the case of hadron propagators, loop diagrams including all tastes are the leading contributions in many thermodynamic quantities.
Therefore, a good control of the taste violation is important in finite temperature QCD.

Another problem which is worrisome in particular in finite temperature QCD is if we can expect scaling properties at finite lattice spacings in spite of non-local interactions.
This matters, e.g.\ when we study scaling properties around a second order chiral transition or near a critical end point of a first order transition line.
I know no theoretical arguments supporting applicability of universality arguments with rooted staggered quarks on finite lattices.

I come back to these issues later.

\subsection{Wilson-type quarks}

A more conservative approach is to adopt Wilson-type lattice quarks with which the existence of the continuum limit is guaranteed for any number of flavors, 
at the price of explicit violation of the chiral symmetry at finite lattice spacings and more computational costs near the massless limit.
In the studies with Wilson-type quarks, we need to keep in mind that the phase structure is not simple at finite lattice spacings due to the explicit violation of the chiral symmetry.
In 1984, Sinya Aoki predicted the existence of a second order phase transition line in the phase diagram to explain massless pions without resorting to chiral symmetry \cite{SAoki}.
Beyond the second order transition line, we have the ``Aoki phase'' in which a parity-flavor symmetry is spontaneously broken. 
Fourteen years later, Sharpe and Singleton pointed out that the chiral violation of Wilson quarks induces an additional term in the chiral effective theory, and the phase structure at finite lattice spacings changes depending on the sign of the corresponding effective coupling $c_2$ \cite{SharpeSingleton}.
Namely, when $c_2$ is positive, we have the second order transition as predicted by Aoki.
But, when $c_2$ is negative, we instead have a first order transition.

\begin{figure}[tbh]
\begin{center}
  \includegraphics[width=0.285\textwidth]{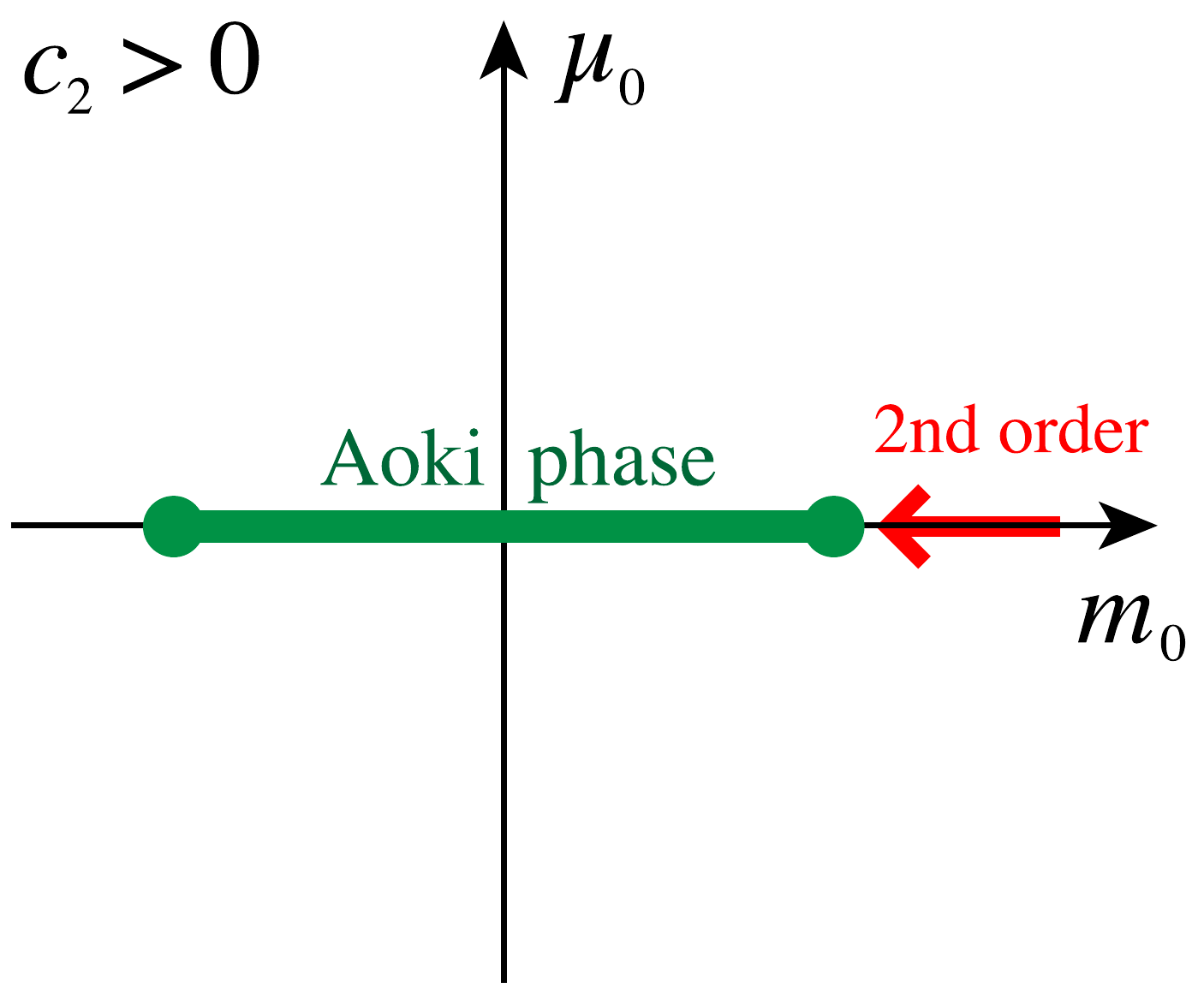}
\hspace{0.5mm}
  \includegraphics[width=0.285\textwidth]{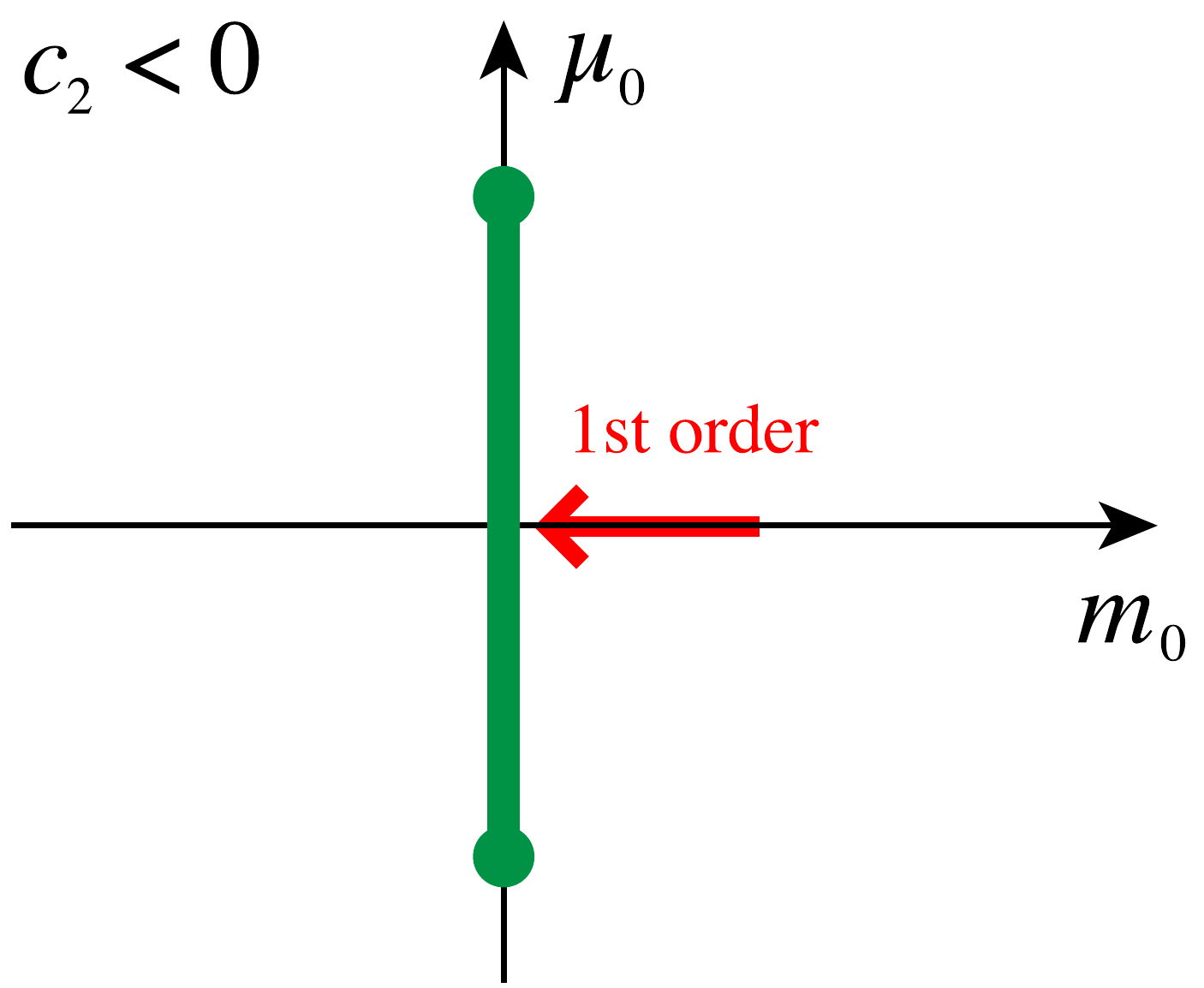}
\hspace{0.5mm}
  \includegraphics[width=0.39\textwidth]{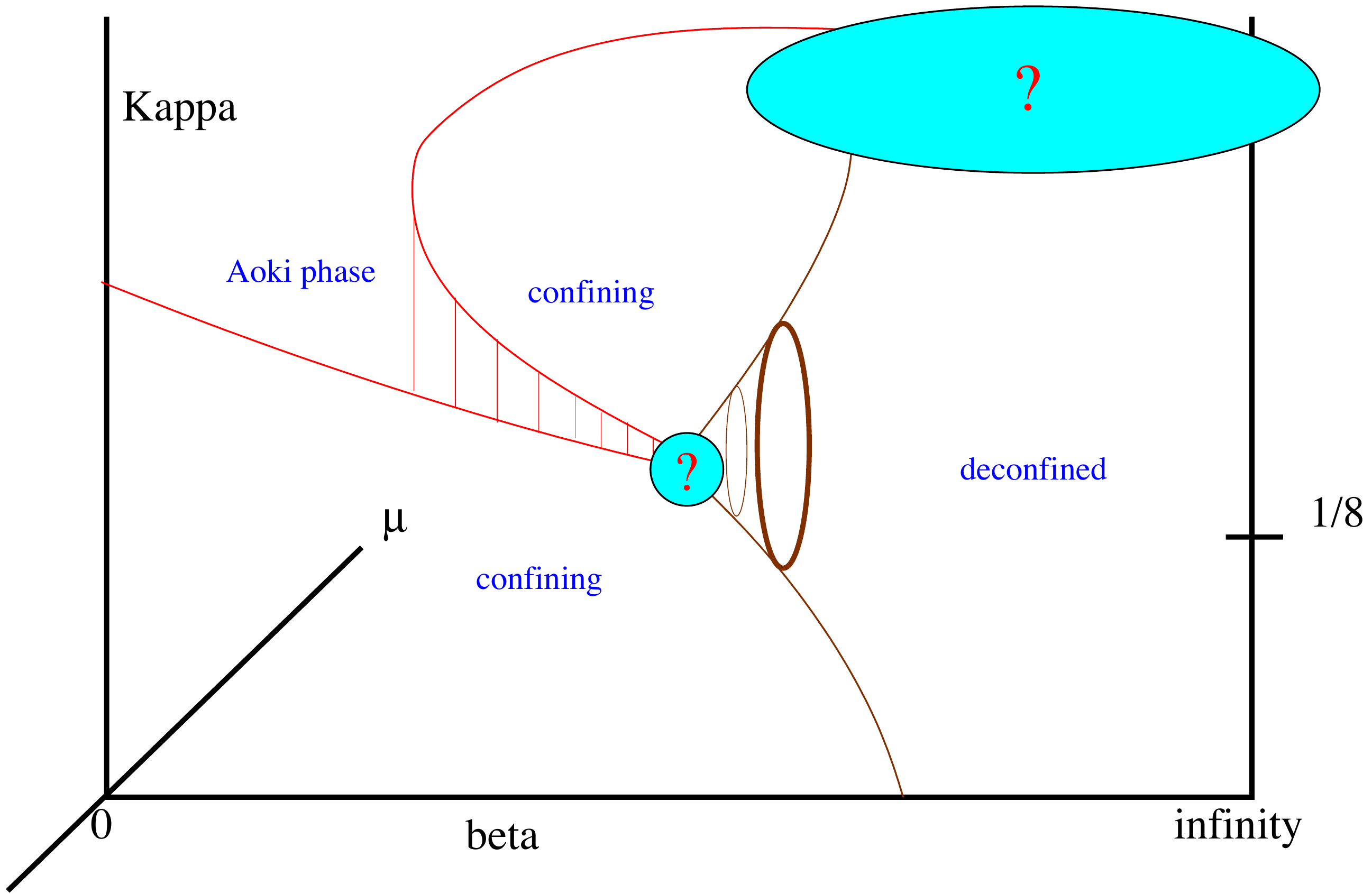}
\end{center}
\vspace{-6mm}
\caption{
{\bf (Left, Center)} 
Phase structure for twisted Wilson-type quarks at finite lattice spacing as a function of the untwisted bare mass parameter $m_0$ and the twisting parameter $\mu_0$, based on a prediction by Ref.~\cite{SharpeWu}.
{\bf (Right)} 
Suggested phase structure for twisted Wilson-type quarks at finite $N_t$ \cite{CreutzPRD76}. 
The vertical axis is the hopping parameter. 
Possible first order cuts in the $\mu$ direction are hidden in the regions marked by ``?''.
}
\label{Fig:tm}
\end{figure}

The phase structure becomes more transparent when we extend the action to include a mass twisting term \cite{tm},
$\displaystyle{
S^{\rm tm}_q = a^4 \sum_x \bar{\psi}(x) \,\left[ D_{\rm W} + m_0 + i\gamma_5\tau_3\mu_0\right]\,\psi(x)
}$,
where $D_{\rm W}$ is the Wilson quark kernel and $\tau_3$ acts in the isospin space.
Sharpe and Wu \cite{SharpeWu} studied the chiral effective theory for twisted quarks and found a phase structure schematically plotted in the left and center panels of Fig.~\ref{Fig:tm}.
The untwisted theories ($\mu_0=0$) are on the horizontal axis.
For the case of $c_2>0$, we have the first order cut corresponding to the Aoki phase on the horizontal axis. 
When we decrease the mass $m_0$ along the horizontal axis, we hit the second order boundary of the Aoki phase, as shown by a red arrow in the figure.
When $c_2<0$, on the other hand, the cut appears along the vertical axis and we encounter a first order transition when we decrease $m_0$. 

Although the sign of $c_2$ should depend on the details of the lattice action, various gauge actions combined with twisted standard Wilson quark action are known to have the first order transition at large $\beta$.
In this case, we have to avoid the first order region to approach the continuum limit.
While the width of the first order region is expected to shrink rapidly towards the continuum limit,  a region with sufficiently light quarks may exist only at small lattice spacings.
On the other hand, the PACS-CS Collaboration have studied QCD with non-perturbatively improved clover quarks coupled to the RG-improved Iwasaki glue, and succeeded in decreasing the light quark mass down to the physical point without encountering a first order transition \cite{PACSCS}.
Therefore, for this combination of the actions, either $c_2>0$ is realized or the first order transition region is located at quark masses lighter than the physical light quark mass.
The phase structure with twisted improved quarks is not well clarified yet \cite{tmC}.

At finite temperature, Creutz \cite{CreutzPRD76} studied a chiral effective model for twisted Wilson quarks and suggested that the deconfinement transition surface in the extended parameter space forms a deformed cone-like shape, as shown in the right panel of Fig.~\ref{Fig:tm}. 
The tmfT Collaboration studied twisted standard Wilson quarks, and obtained signals consistent with the predicted cone-shaped deconfinement transition surface \cite{tmftPRD80,ZeidlewiczLat10}.
Further studies are needed to clarify the full phase structure for twisted Wilson-type quarks.

\subsection{Lattice chiral quarks}

The most attractive but currently the most expensive way to push forward is to adopt a chiral lattice quark action.
Domain-wall (DW) quarks \cite{DW} are formulated on a 5-dimensional space-time and designed such that the chiral symmetry is recovered at finite lattice spacings in the limit $L_s \rightarrow \infty$  where $L_s$ is the lattice size in the 5th dimension.
In actual simulations, however, we have to keep $L_s$ finite.
This causes a residual chiral violation, which is conventionally measured by the magnitude of the additive renormalization term $m_{\rm res}$ to quark masses near the continuum limit.
$m_{\rm res}$ is called the residual quark mass and is known to decrease exponentially with $L_s$ for small $L_s$, but then levels off into a much slower decrease proportional to $1/L_s$ due to the mobility edge phenomenon \cite{Mres}.
The requirement of large computational resources 
forces us to work on lattices coarser than other lattice quarks and with small $L_s$.
Because DW quarks require smooth gauge configurations, coarse lattices may introduce serious chiral violations.
A study of finite temperature QCD with DW quarks was started about 10 years ago for $N_F=2$ on a coarse lattice ($8^3\times4$) with $L_s \approx 8$ \cite{ChenPRD64}.
Although the chiral violations seem to be controled by $m_{\rm res}$, the values of $m_{\rm res}$ were too large to discuss physical properties.

\FIGURE{
 \includegraphics[width=0.435\textwidth]{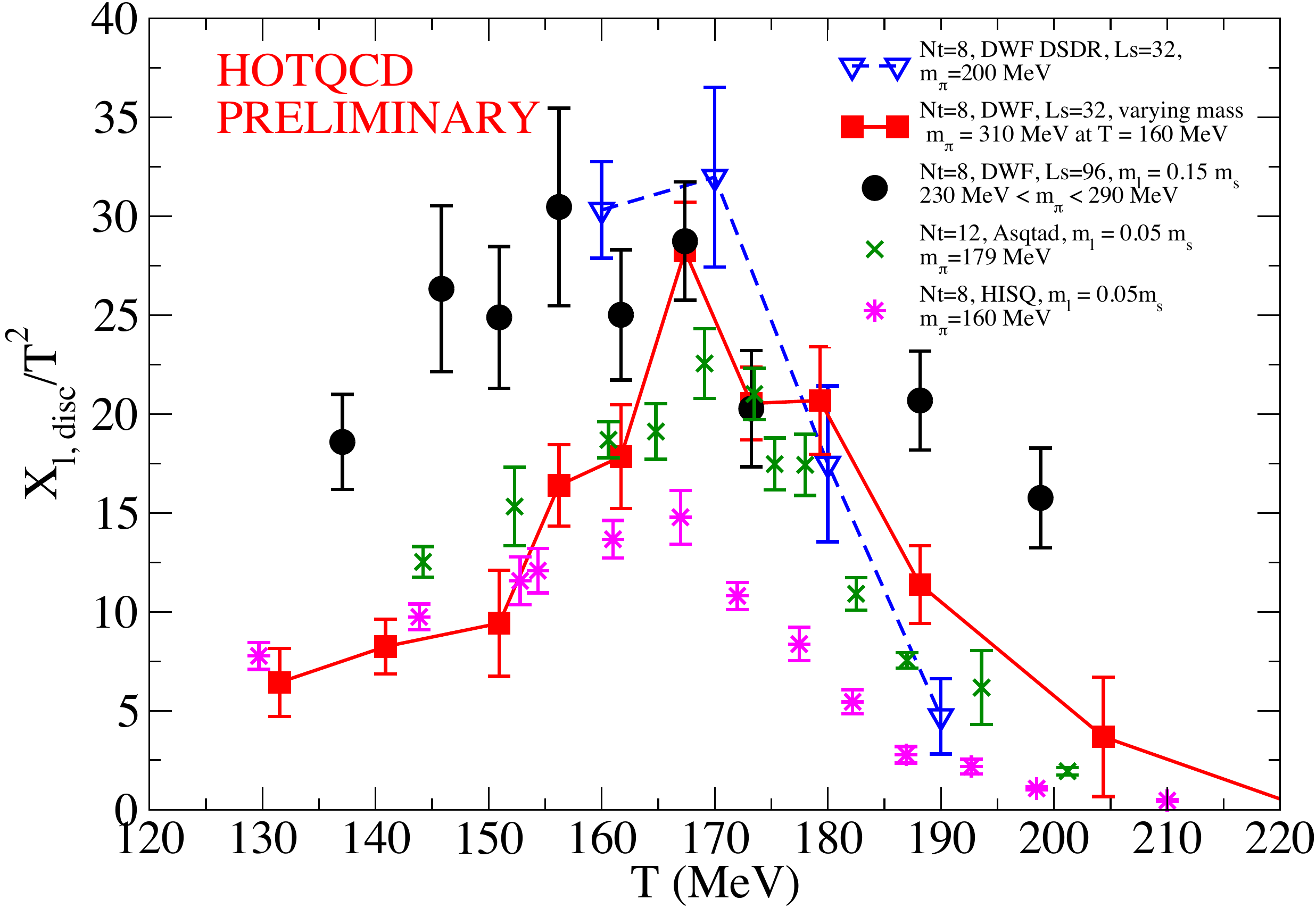}
\vspace{-2.5mm}
\caption{Disconnected chiral susceptibility obtained with 2+1 flavors of DW quarks 
(red square) \cite{ChengPRD81}, 
compared with the results of asqtad and HISQ quarks \cite{BazavovSoeldnerLat10}
Preliminary results for $L_s=96$ (black circles) and 
with an improved gauge action 
(blue downward triangles) \cite{ChengLat10} are also shown.}
\label{Fig:HotQCDDW}
}

This year, a new study of 2+1 flavor QCD adopting the RG-improved Iwasaki gauge action and much larger $L_s\approx32$ on a finer lattice  ($16^3\times8$) has been published  \cite{ChengPRD81}.
Both the improvement of gauge action and the adoption of a larger $L_s$ are effective to reduce $m_{\rm res}$.
Accordingly, their results for chiral susceptibilities around the thermal transition point show an effective quark mass dependence qualitatively consistent with expectations.
However, even with these improvements, the resulting $m_{\rm res}$ turned out to be not small enough:
$m_{\rm res} \sim 0.008$ is larger than the bare light quark mass $m_l  = 0.003$ needed to achieve the effective light to strange quark mass ratio $(m_l+m_{\rm res})/(m_s+m_{\rm res}) \approx 0.025$.
The HotQCD Collaboration is now extending the study adopting an even larger $L_s$ of 96, and also with an improved action \cite{Renfrew} dedicated for DW quarks.
See Fig.~\ref{Fig:HotQCDDW} \cite{ChengLat10}.

Finally, Cossu reported at the conference the status of a finite temperature study that JLQCD collaboration has started using an overlap quark at a fixed topology \cite{CossuLat10}.

\section{Phase structure of QCD at finite temperatures}
\label{sec:PhaseStructure}

\begin{figure}[tbh]
\vspace*{-2mm}
\begin{center}
  \includegraphics[width=0.4\textwidth]{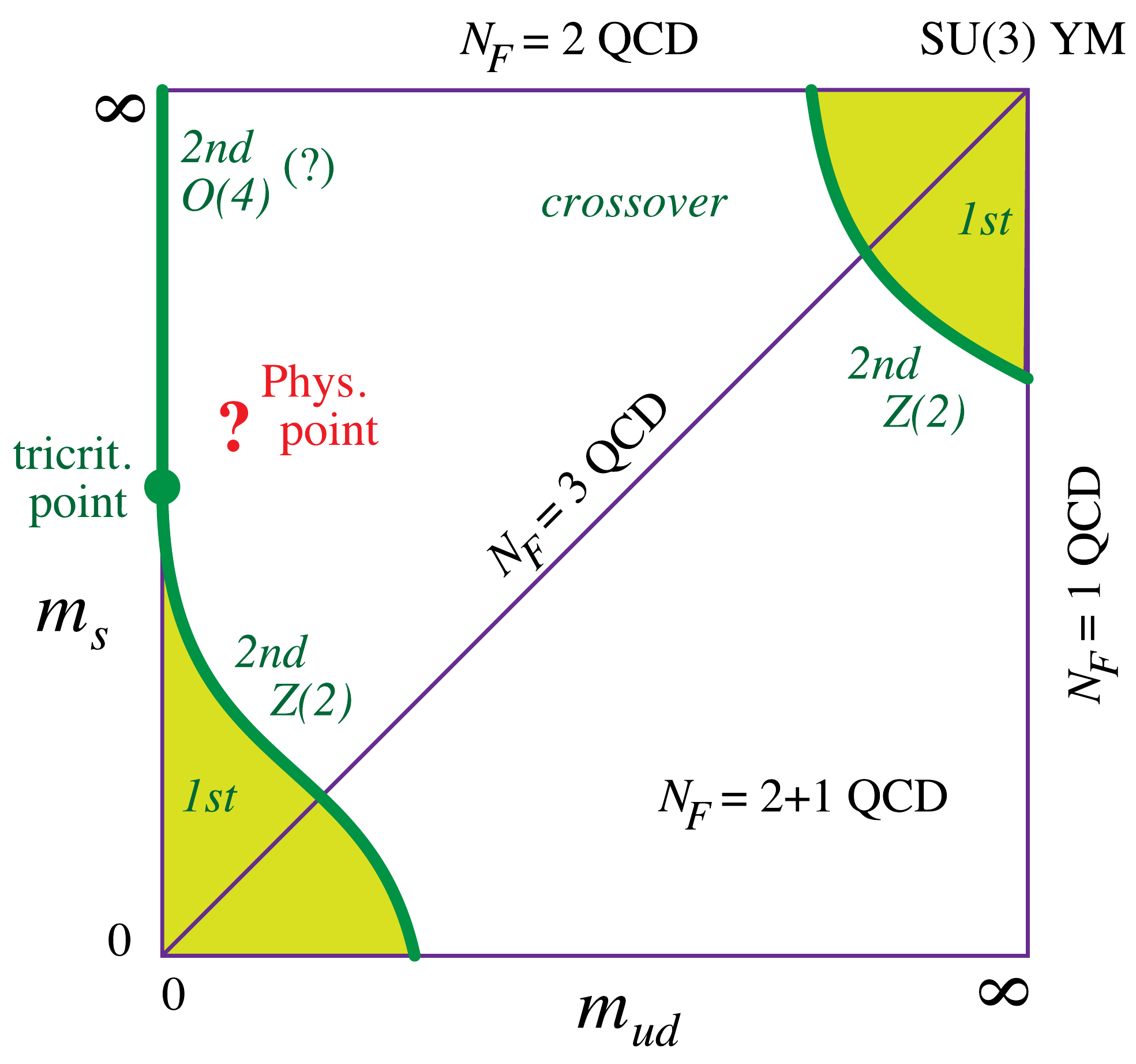}
  \hspace{7mm}
  \includegraphics[width=0.4\textwidth]{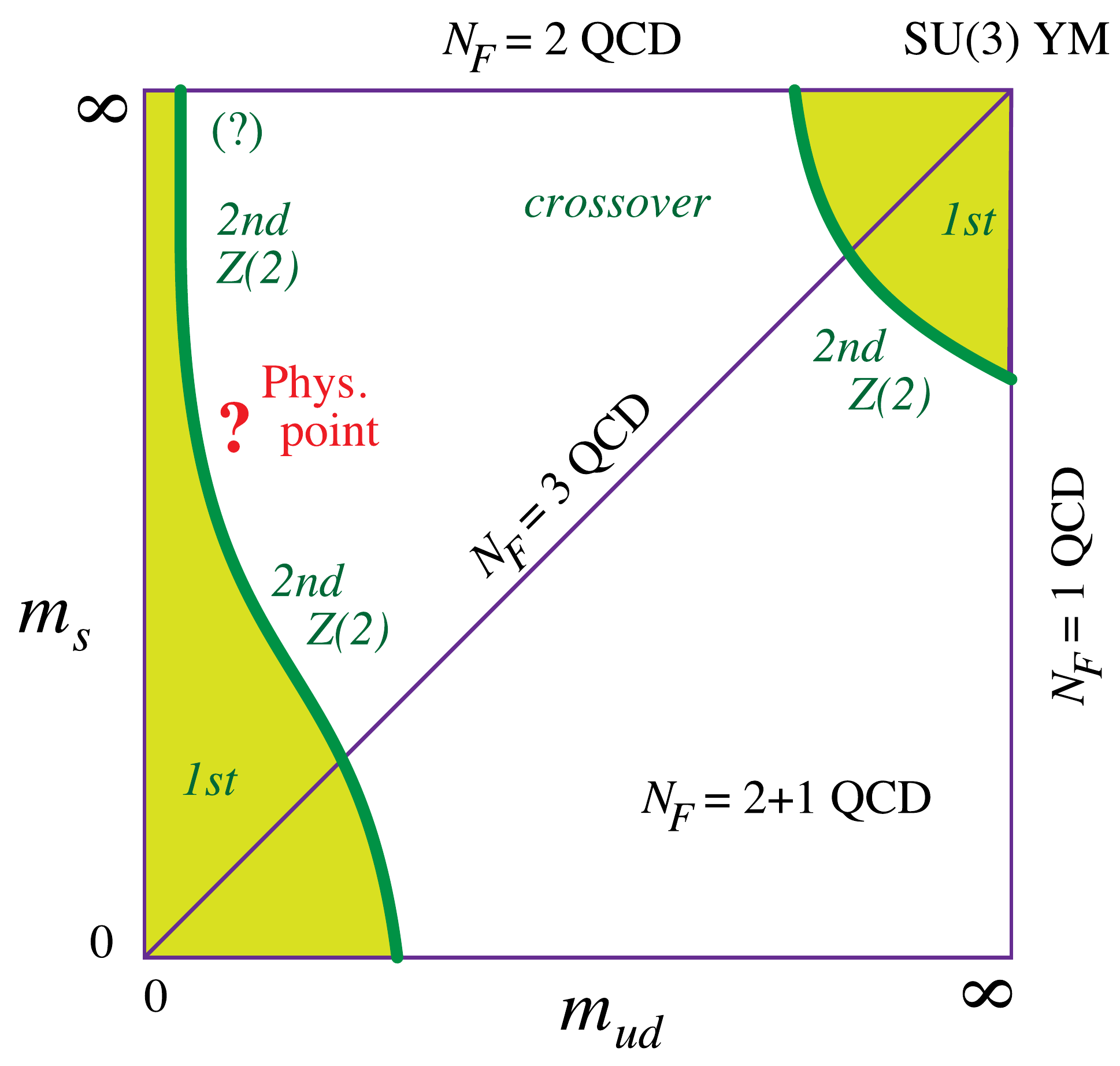}
\end{center}
\vspace{-6mm}
\caption{Order of the finite temperature transition in 2+1 flavor QCD as a function of the degenerate $u$ and $d$ quark mass $m_{ud}$ and the $s$ quark mass $m_s$.
{\bf (Left)} 
The standard scenario with the second order chiral transition for two-flavor QCD. 
{\bf (Right)} 
An alternative scenario when the two-flavor chiral transition is first order.}
\label{Fig:Nf21PhaseDiagram}
\end{figure}

Figure~\ref{Fig:Nf21PhaseDiagram} summarizes the current wisdom about the order of the finite temperature transition in 2+1 flavor QCD as a function of the quark masses,
based on the studies of lattice simulations and effective models.
When the $u$, $d$ and $s$ quarks are all sufficiently light or sufficiently heavy, we expect the transition to be of first order.
At intermediate values of the quark masses, the ``transition'' will be actually an analytic crossover.
On the boundaries of the first order regions, we expect second order transitions.
Associated critical scaling around there will have characteristics universal to the corresponding effective models.

Simulations using improved staggered quarks suggest that the physical point, identified by the pNG pion mass and other hadron masses, is located in the crossover region (see e.g.\ \cite{WB_Nature}). 
Now, the main challenges are to compute the detailed properties of the transition at the physical point with a high precision. 
The transition temperature $T_c$ is one of the most important quantities for experimental investigations of QGP.
Estimation of $T_c$ in 2+1 flavor QCD has been made based on large-scale simulations using various improved staggered quarks. 
However, there has been a big discrepancy in the values of $T_c$ among different groups for more than five years.
This year, the main part of the discrepancy has been removed.

The nature of the transition in the chiral limit of two-flavor QCD (the upper left edge of the figure) has significant implications for the nature of the transition at the physical point too.
The left panel of Fig.~\ref{Fig:Nf21PhaseDiagram} summarizes the standard scenario in which the chiral transition of two-flavor QCD is second order in the universality class of the O(4) Heisenberg model \cite{PisarskyWilczek}.
In this case, because the chiral transition of three-flavor QCD is of first order, we have a tricritical point on the left edge of the figure ($m_{ud}=0$) where the order of the transition changes from the second order to the first order. 
Depending on the location of the tricritical point relative to the physical point, the universality class dominating the parameter dependence around the physical point will be different.
The right panel of Fig.~\ref{Fig:Nf21PhaseDiagram} shows an alternative scenario in which the chiral transition of two-flavor QCD is first order.
In this case, we have no tricritical point and thus no regions for the O(4) universality class.
A distinction between the two scenarios is important for studies at finite densities too.
Although the majority view the standard scenario as more probable, 
the nature of the two-flavor chiral transition was not fully fixed.
This year, we had some advances. 

In this section, I discuss these developments. 

\subsection{Transition temperature}
\label{sec:Tc}

In 2005, the MILC Collaboration obtained $T_c=169(12)(4)$ MeV in the combined chiral and continuum limit from a measurement of a chiral susceptibility in 2+1 flavor QCD with asqtad quarks and the one-loop Symanzik glues on $N_t=4$--8 lattices \cite{MilcPRD71}, where the scale was set by $r_1$ and the O(4) critical exponent was adopted in the chiral extrapolation.
In 2006, the Wuppertal-Budapest Collaboration has published their values based on a study of the 2+1 flavor QCD with a stout-link improved staggered quarks coupled to the tree-level Symanzik glues \cite{WB_Tc1}.
Carrying out a chiral extrapolation to the physical point and a continuum extrapolation using $N_t=6$--10 lattices, they found 
$T_c = 151(3)(3)$ MeV from the susceptibility of the chiral condensate, 
$T_c = 175(2)(4)$ MeV from the $s$ quark number susceptibility, 
and $T_c = 176(3)(4)$ MeV from the inflection point of the Polyakov-loop,
where the scale was set by $f_K$.
In the same year, Cheng et al.\  (let me call them as ``BNL-Bielefeld Collaboration'') 
published $T_c = 192(7)(4)$ MeV at the physical point in the continuum limit,
based on their study on $N_t=4$ and 6 lattices using the p4fat3 staggered quark action and tree-level Symanzik gauge action \cite{BNLB_Tc1}.
Their $T_c$ is an average of estimates from the Polyakov-loop susceptibility and the chiral susceptibility, and the systematic error includes the difference between them.
Later in 2008, the HotQCD Collaboration (which is a merger of the BNL-Bielefeld Collaboration and the MILC Collaboration) compared the results of p4 and asqtad actions, and found that the two staggered quarks lead to roughly consistent results for the location of the transition on finite lattices \cite{BNLB_EOSa,BNLB_EOSt8}.

\begin{figure}[tbh]
\vspace{1mm}
\begin{center}
  \includegraphics[width=0.275\textwidth]{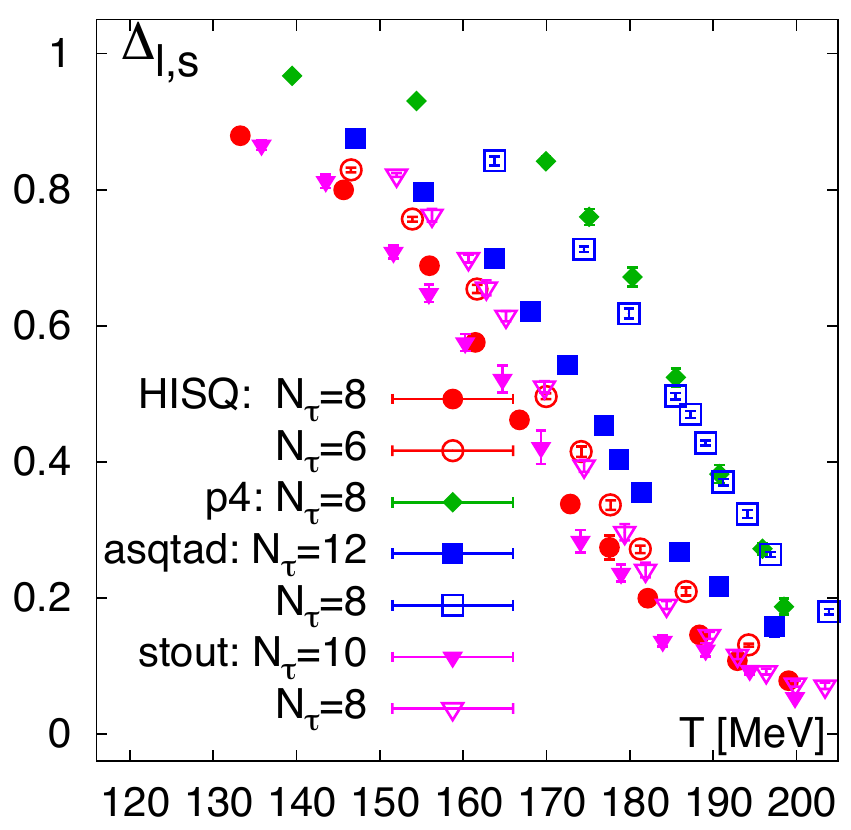}
\hspace{2mm}
  \includegraphics[width=0.335\textwidth]{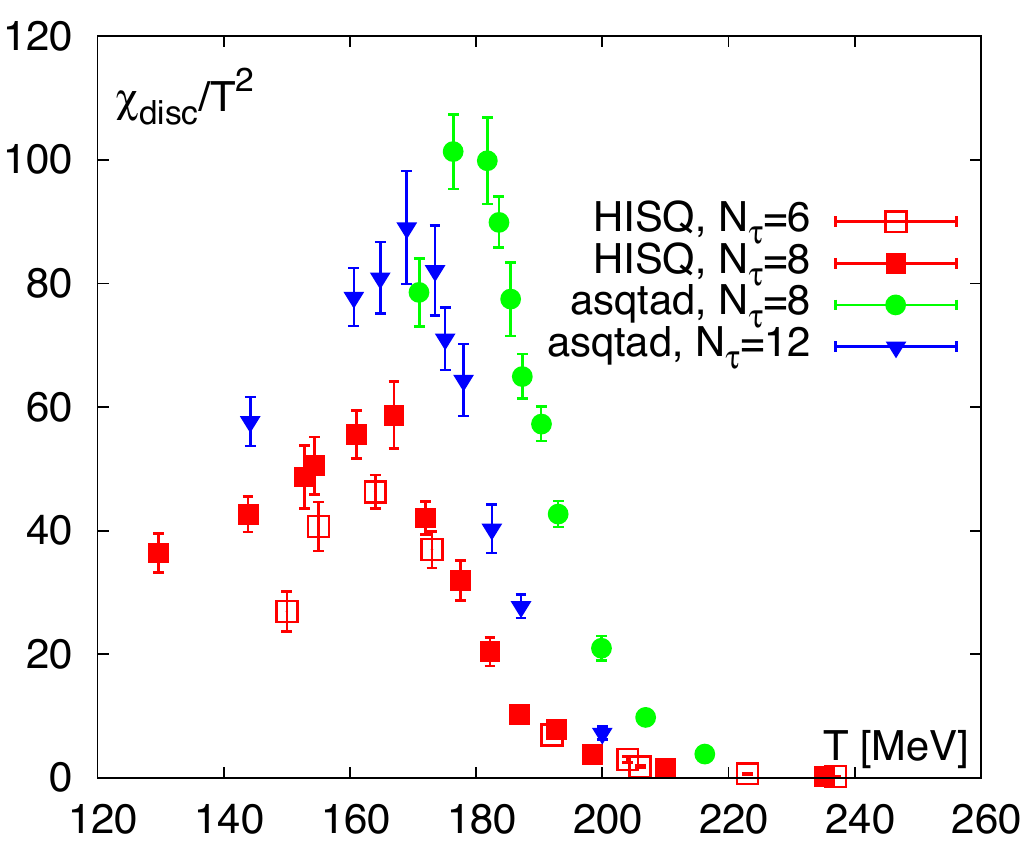}
\hspace{1.5mm}
  \includegraphics[width=0.32\textwidth]{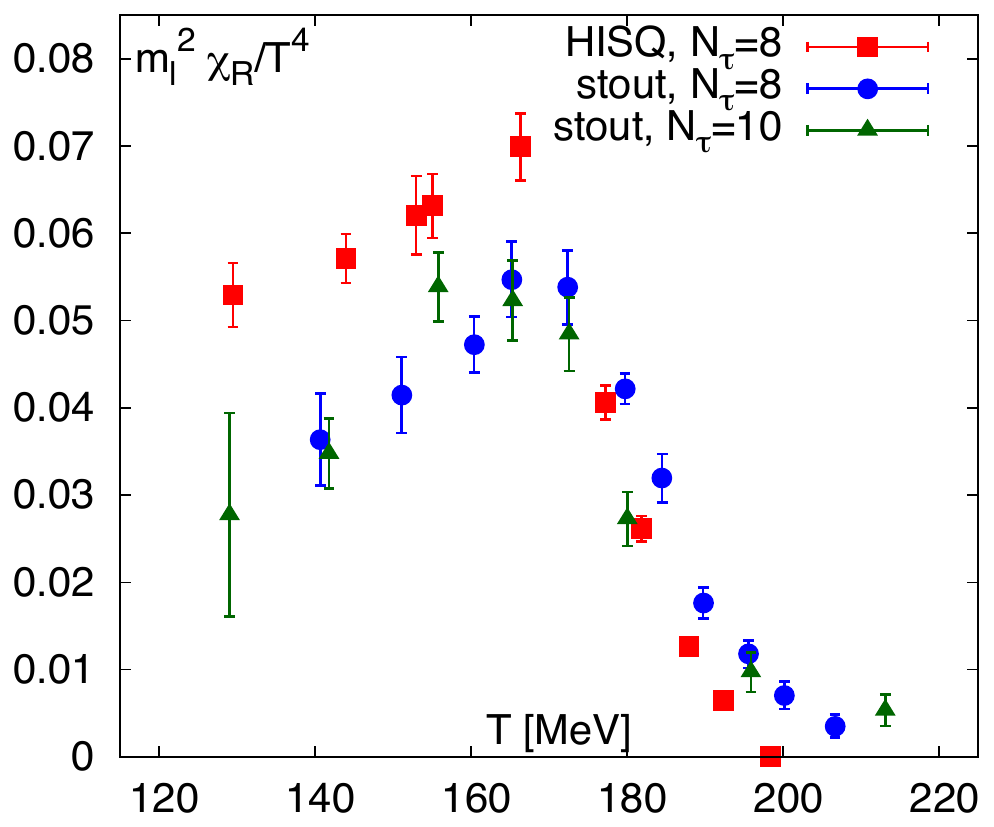}
\end{center}
\vspace{-4.5mm}
\caption{
Chiral observables in 2+1 flavor QCD with improved staggered quarks.
The HISQ, p4 and asqtad data are obtained at the bare quark mass ratio $m_{ud}/m_s = 0.05$ with $m_s$ around the physical value, using a scale fixed by $r_0$ .
The physical point is estimated to be $m_{ud}/m_s \approx 0.037$.
The stout data, which is obtained at the physical point using a $f_K$ scale \cite{WB_Tc3}, is shifted in $T$ to correct the difference in the scale setting.
Note that this procedure causes a slight deviation from the physical point at finite lattice spacings.
{\bf (Left)}
Subtracted chiral condensate (difference of scaled light quark and $s$ quark condensates to remove a divergent renormalization factor) from the HISQ, p4, asqtad and stout actions \cite{BazavovSoeldnerLat10}.
{\bf (Center)}
Comparison of HISQ and asqtad results for the disconnected part of the chiral susceptibility \cite{BazavovSoeldnerLat10}.
{\bf (Right)} 
Comparison of HISQ and stout results for the renormalized chiral susceptibility in which the $T=0$ contribution is subtracted \cite{BazavovPetrecky}. 
}
\label{Fig:Tc1}
\end{figure}

Because we have an analytic crossover instead of a singular phase transition, the value of $T_c$ may depend on the choice of observable to define it.
Details of the analyses, such as the scale setting and the definition of the line of constant physics, are different among the two groups too. 
However, it was shown that the discrepancy exists even when we adopt the same observable and the same scale convention \cite{WB_Tc2}. See the left panel of Fig.~\ref{Fig:Tc1}.
We find that the whole temperature dependence is shifted depending on the choice of the action.

Both groups have pushed forward their studies to clarify the origin of the discrepancy.
The Wuppertal-Budapest Collaboration extended the study to finer lattices ($N_t$ up to 16) and concluded that their values remain essentially unchanged \cite{WB_Tc2,WB_Tc3}.
Their final values are 
$T_c = 147(2)(3)$ MeV from the susceptibility of the chiral condensate 
and $T_c = 165(2)(4)$ MeV from the $s$ quark number susceptibility \cite{WB_Tc3}.
On the other hand, the HotQCD Collaboration noted that the transition temperature from p4 and asqtad quarks show rather rapid shift of $-5$ to $-7$ MeV on finite lattices when the lattice spacing is decreased from $N_t=6$ to 8 \cite{BNLB_EOSt8},
and another shift of about $-5$ MeV when the bare light quark mass is decreased from $m_{ud}/m_s = 0.1$ to 0.05, that corresponds to a decrease of the pNG pion mass from about 220 MeV to 160 MeV \cite{BNLB_EOSp}.
At the conference, they presented a preliminary result $T_c=164(6)$ MeV at the physical point in the continuum limit, 
from their study of disconnected chiral susceptibility using asqtad quarks on lattices up to $N_t=12$
\cite{BazavovSoeldnerLat10}.
At the same time, the Hot QCD Collaboration has presented first results of their project on QCD thermodynamics adopting a new improved staggered quark action, the ``highly improved staggered quark (HISQ)'' action \cite{BazavovSoeldnerLat10,BazavovPetrecky}.
A comparison of these actions is shown in Fig.~\ref{Fig:Tc1}.

Consulting the recent values of $T_c$ and the location of the peaks in the chiral susceptibilities shown in Fig.~\ref{Fig:Tc1}, we find that the discrepancy in $T_c$ is now mostly removed: All the actions converge towards $T_c \approx145$--165 MeV.

What was the origin of the discrepancy?
The main difference among the improved staggered quark actions seems to be the magnitude of the remaining taste violation:
When we adjust the bare quark mass to have $m_\pi^{\rm pNG} \approx 135$ MeV at lattice spacings corresponding to $T\approx170$ MeV on $N_t=8$ lattices, the masses for the heavy pions are around 400--600 MeV (asqtad), 300--500 MeV (stout) and 200--400 MeV (HISQ), respectively. 
The p4 is slightly worse than the asqtad.
Therefore, the p4 and asqtad quarks suffer from severer lattice artifacts than the stout and HISQ quarks.
Note that the simulations with stout quarks were performed on finer lattices.
On an $N_t=12$ lattice, the heavy pion masses with the stout quark action will be around 200--350 MeV at $T\approx170$ MeV.

\FIGURE{
  \includegraphics[width=0.385\textwidth]{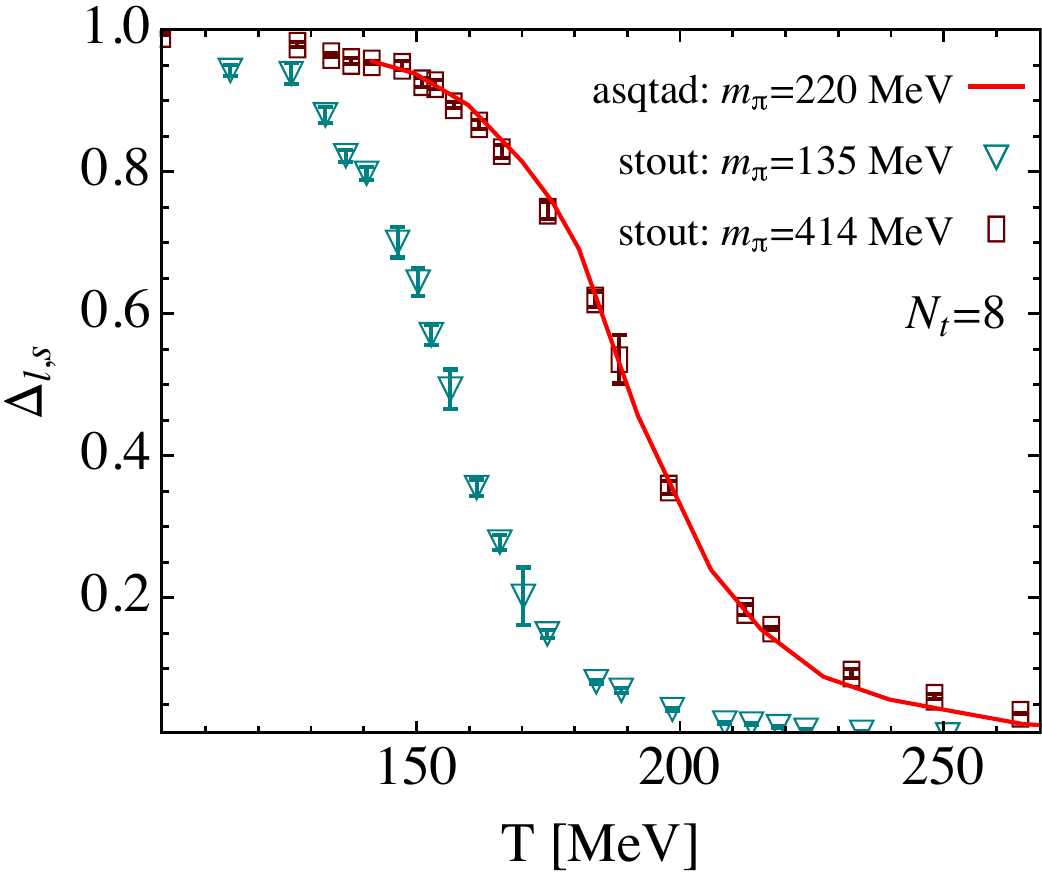}
\vspace{-1.5mm}
\caption{
Subtracted chiral condensate with stout and asqtad quarks at $N_t=8$ \cite{WB_Tc3}. 
}
\label{Fig:Tc2}
}

The overall effects of light and heavy pions in loop diagrams may be measured by a root mean squared (RMS) mass of pions $m_\pi^{\rm RMS}$ \cite{DeTarLat08}.
In Fig.~\ref{Fig:Tc2}, the Wuppertal-Budapest Collaboration demonstrated that the asqtad data can be reproduced by the stout quark action by adjusting $m_\pi^{\rm RMS}$ \cite{WB_Tc3}.
The open rectangles are results of stout quark action at $m_\pi^{\rm pNG}=414$ MeV \cite{WB_Tc3},  and the red curve is the result of asqtad quark action at $m_\pi^{\rm pNG}=220$ MeV \cite{BNLB_EOSt8}.
The stout quark mass for open rectangle is adjusted to reproduce $m_\pi^{\rm RMS}=587$ MeV of the asqtad data at $T\approx 135$ MeV on $N_t=8$.
According to the larger taste violation with the asqtad action, the asqtad data at a given $m_\pi^{\rm RMS}$ corresponds to the stout data at a larger $m_\pi^{\rm pNG}$.
This explains the larger lattice artifacts in the p4 and asqtad results at $N_t$ \lsim\ 8.

While the major part of the discrepancy is explained by the taste violation, small disagreements remain.
In particular, the sensitivity of the determination of $T_c$ on the operator, observed with the stout quark, is not confirmed by other quarks yet.

In all of these studies, the physical point is identified by the pNG pion mass.
The success of the RMS pion mass, however, suggests that it is more appropriate to consult the RMS hadron masses to judge the location of the physical point for thermodynamic quantities.
This will make the physical point closer to the chiral limit and the values of $T_c$ even smaller on finite lattices. 

It is highly desirable to confirm the results with other types of lattice quarks.
Unfortunately, the current status with Wilson-type quarks is quantitatively much behind.
Recent updates of $T_c$ in two-flavor QCD are given in Refs.~\cite{ZeidlewiczLat10,Bornyakov09102392,WHOT_PRD82}.
The first rough estimate of $T_c$ with 2+1 flavors of DW quarks is given in \cite{ChengPRD81}.

\subsection{Chiral scaling}
\label{sec:ChiralScaling}

At $T=0$, QCD in the chiral limit has the global chiral-flavor symmetry ${\rm U}(N_F)_{\rm R} \times {\rm U}(N_F)_{\rm L}$ naively, where $N_F$ is the number of flavors.
Among them, U$(1)_{\rm A}$ is explicitly violated by the anomaly, 
and  the remaining symmetry is dynamically broken down as
$
{\rm SU}(N_F)_{\rm R} \times {\rm SU}(N_F)_{\rm L} \times {\rm U}(1)_{\rm V}
\;\longrightarrow\;
{\rm SU}(N_F)_{\rm V} \times {\rm U}(1)_{\rm V}
$. 
At sufficiently high temperatures, the broken symmetry will be restored.
From a study of effective Ginsburg-Landau theory respecting these symmetries \cite{PisarskyWilczek}, 
the chiral restoration transition is predicted to be of first order for $N_F \ge 3$.
For the case of $N_F=2$, the transition is predicted to be either second order in the universality class of O(4), or first order if the anomaly is weak around the transition point.
Clarification of the $N_F=2$ chiral transition is important for understanding the nature of the transition at the physical point, as summarized in Fig.~\ref{Fig:Nf21PhaseDiagram}.
Here, scaling analysis is a powerful tool to discriminate the order of phase transitions.

\subsubsection{O(4) scaling with Wilson-type quarks}

\begin{figure}[tbh]
\begin{center}
  \includegraphics[width=0.32\textwidth]{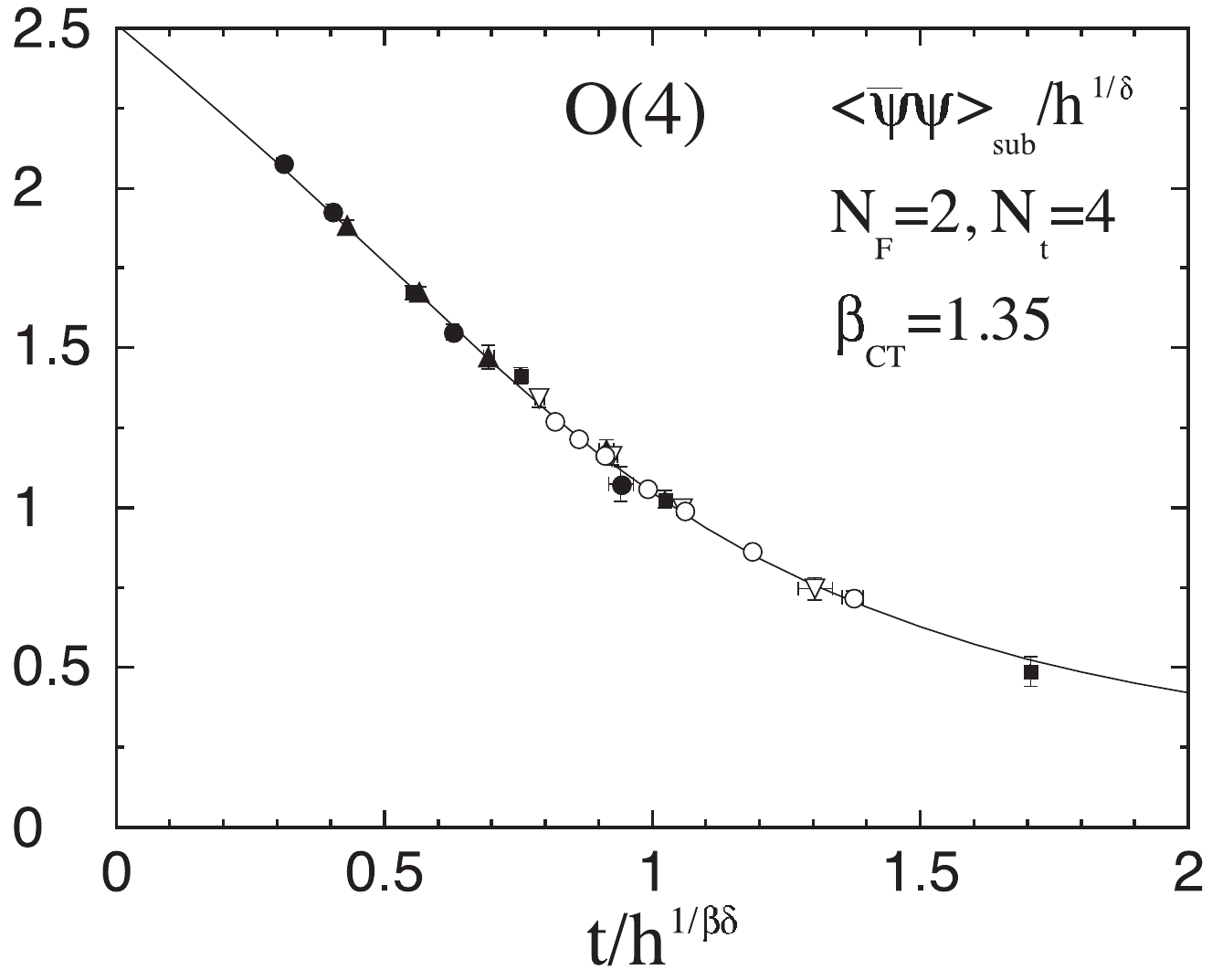}
\hspace{0.5mm}
  \includegraphics[width=0.32\textwidth]{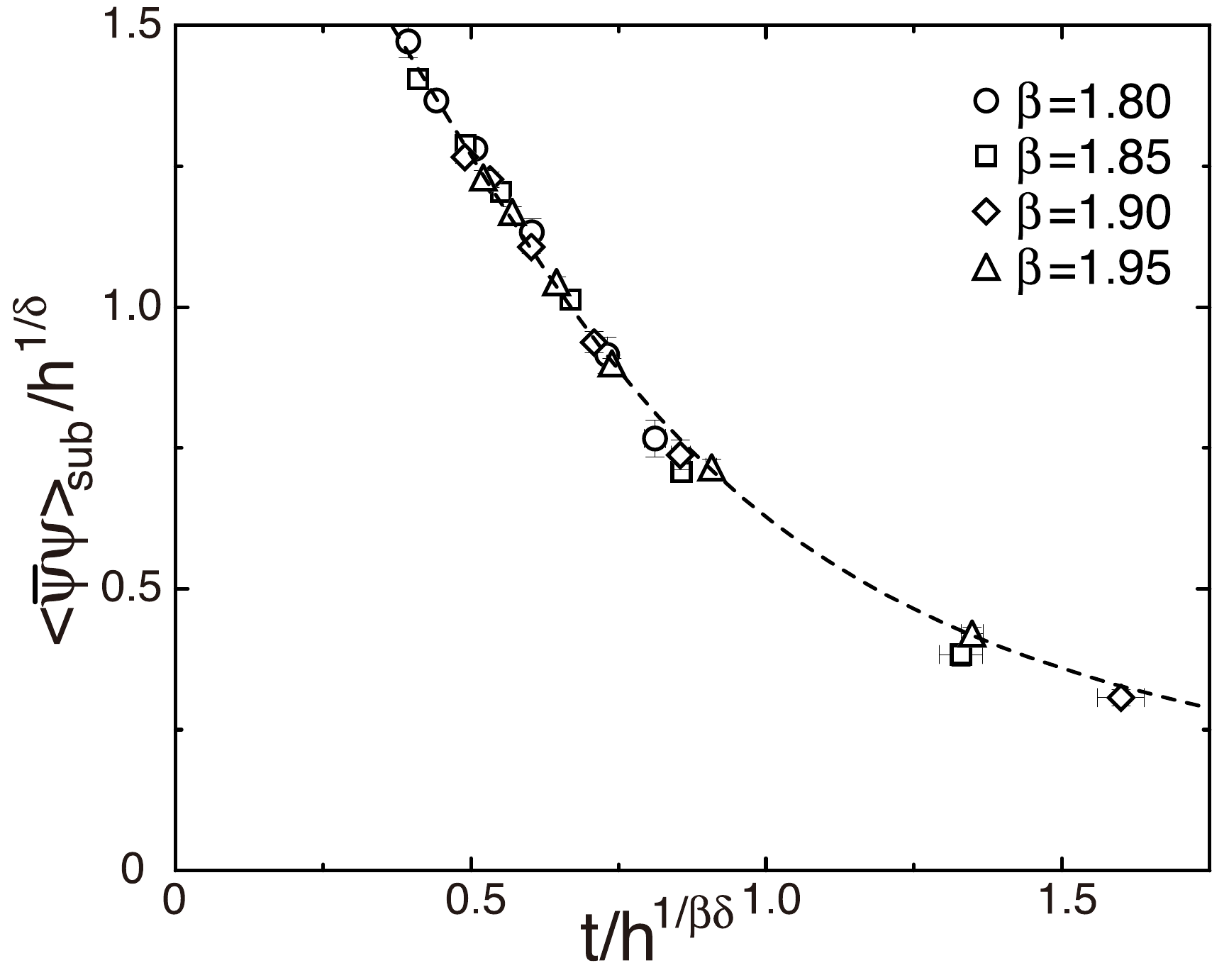}
\hspace{0.5mm}
  \includegraphics[width=0.315\textwidth]{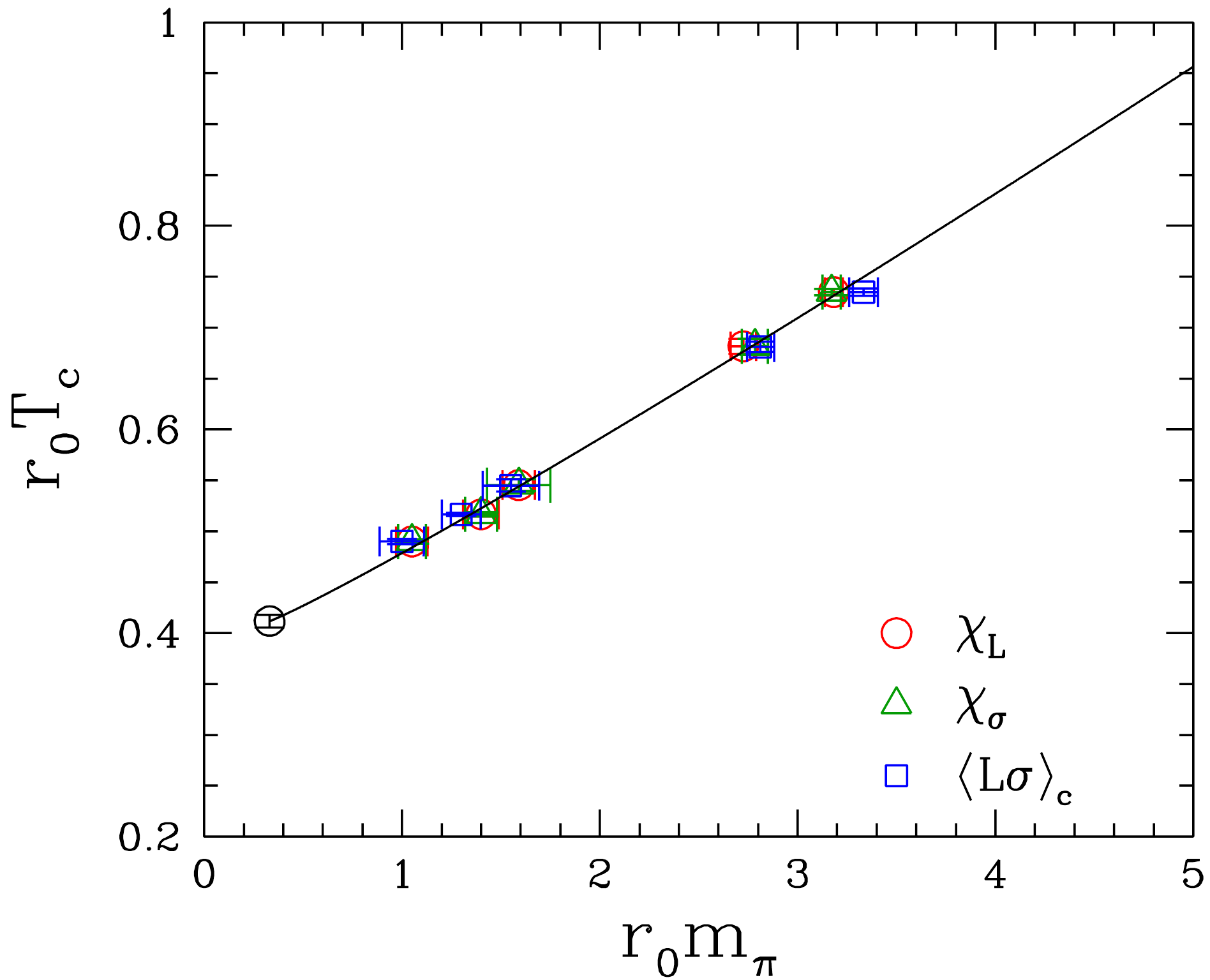}
\end{center}
\vspace{-6mm}
\caption{
O(4) scaling fits with in two-flavor QCD with Wilson-type quarks.
{\bf (Left)} 
Chiral condensate with the unimproved Wilson quark action and the Iwasaki gauge action \cite{IwasakiPRL78}.
The chiral condensate is rescaled using O(4) critical exponents, and the curve is the scaling function obtained in an O(4) spin model.
{\bf (Center)}
Rescaled chiral condensate with a MF-improved clover quark action and the Iwasaki gauge action \cite{CppacsPRD63}.
The dashed curve is the scaling function obtained in an O(4) spin model.
{\bf (Right)} 
O(4) fit to the pseudocritical temperature $T_c$ with a NP-improved clover quark action and the unimproved plaquette gauge action \cite{Bornyakov09102392}. 
}
\label{Fig:O4W}
\end{figure}

With Wilson-type quarks, the O(4) scaling expected for the case of a second order chiral transition has been observed  with various combinations of Wilson-type quark actions and gauge actions, since the early years of the investigations. 
Although the chiral symmetry is explicitly violated on the lattice due to the Wilson term, when the additive renormalizations are correctly performed
such that the chiral violations are subtracted out and the chiral Ward identities are satisfied in the continuum limit, 
the resulting subtracted observables will show the correct chiral behaviors near the continuum limit \cite{Bochicchio}.

Adopting a chiral condensate defined by the axial vector Ward identity \cite{IwasakiPRL78},
the QCD data are shown to reproduce the scaling function of the O(4) Heisenberg model, with both unimproved and clover-improved Wilson quark actions \cite{IwasakiPRL78,CppacsPRD63,EjiriLat10}.
See the left and center panels of Fig.~\ref{Fig:O4W}.
As shown in the right panel of the same figure, a recent study on finer lattices with slightly lighter quark masses supports the O(4) behavior too  \cite{Bornyakov09102392}.
These results, together with several observations made in these papers, suggest that the transition is continuous in the chiral limit.
However, the quark masses studied are quite heavy yet ($m_\pi$ \gsim\ 400 MeV).
Confirmation with lighter quarks on a fine lattice is indispensable.
At the conference, Brandt reported the status of their study towards this direction \cite{MainzFrankfurtLat10}.

\subsubsection{Chiral scaling with staggered-type quarks}

Scaling studies with staggered-type quarks have caveats.
$N_F\ne4$ is realized through the fourth root procedure discussed in Sect.~\ref{sec:KS}.
Therefore, the symmetry of the system at finite lattice spacings is that of the original four-taste theory.
In the chiral limit on finite lattices, we thus have the O(2) [$=$ U(1)] symmetry for any number of flavors.

This means that, when the chiral transition is of second order for $N_F=2$, we should expect scaling properties in the universality class of O(2) spin models, instead of the O(4).
Here, we have assumed that the non-locality of rooted staggered quark actions does not invalidate the universality arguments at finite lattice spacings too. 
In any case, the O(4) universality may appear only when we take the continuum limit prior to the chiral extrapolation.
We may, however, expect that when the chiral transition is continuous on finite lattices, it remains so in the continuum limit.
In this sense, examination of O(2) is important, with the caveats mentioned above.

Numerically, it is not very easy to distinguish between O(4) and O(2) scalings with the current accuracy of lattice simulations because the critical exponents and the scaling functions in both theories are almost identical.

\begin{figure}[tbh]
\vspace{1mm}
\begin{center}
  \includegraphics[width=0.40\textwidth]{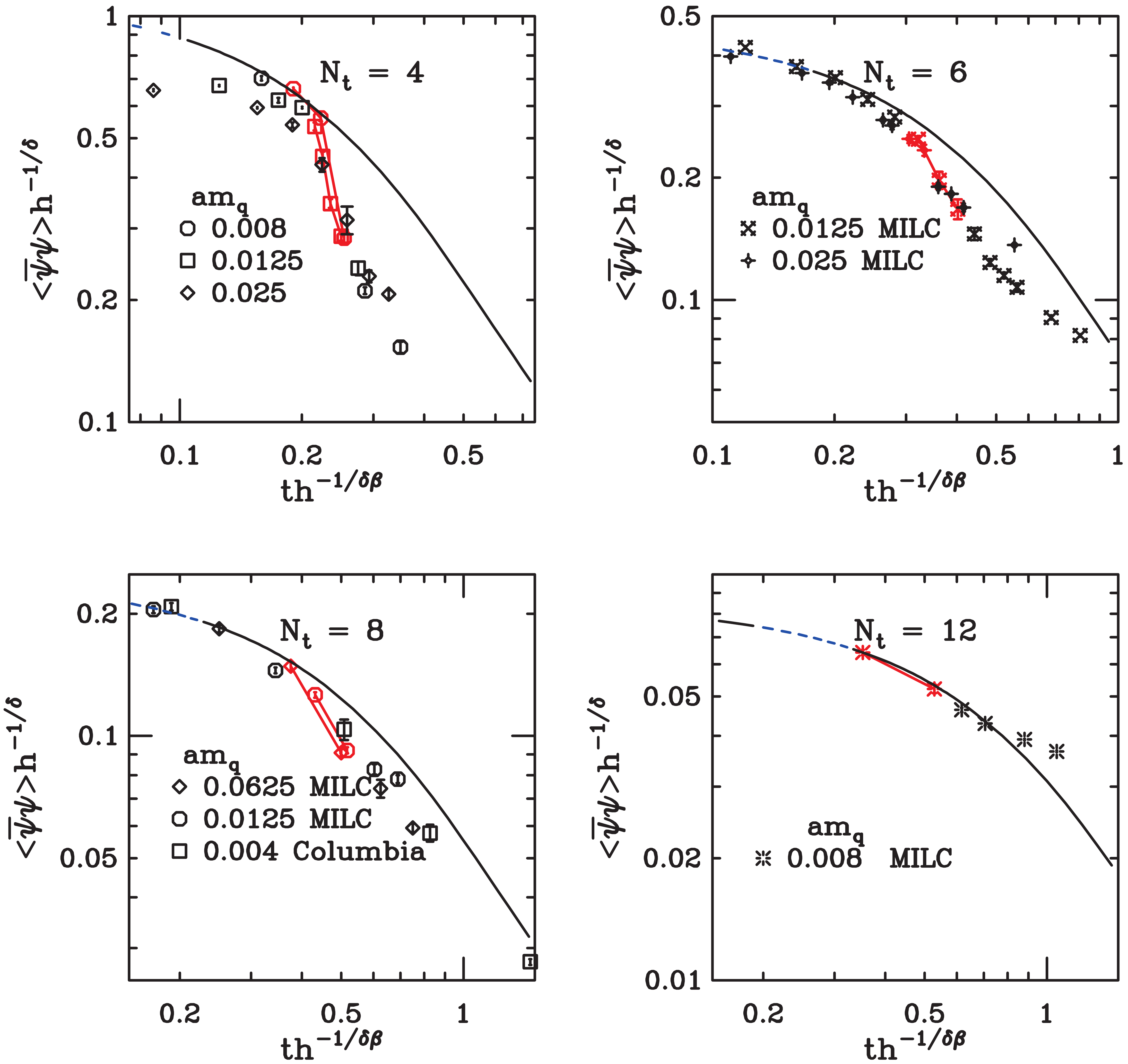}
\hspace{10mm}
  \includegraphics[width=0.37\textwidth]{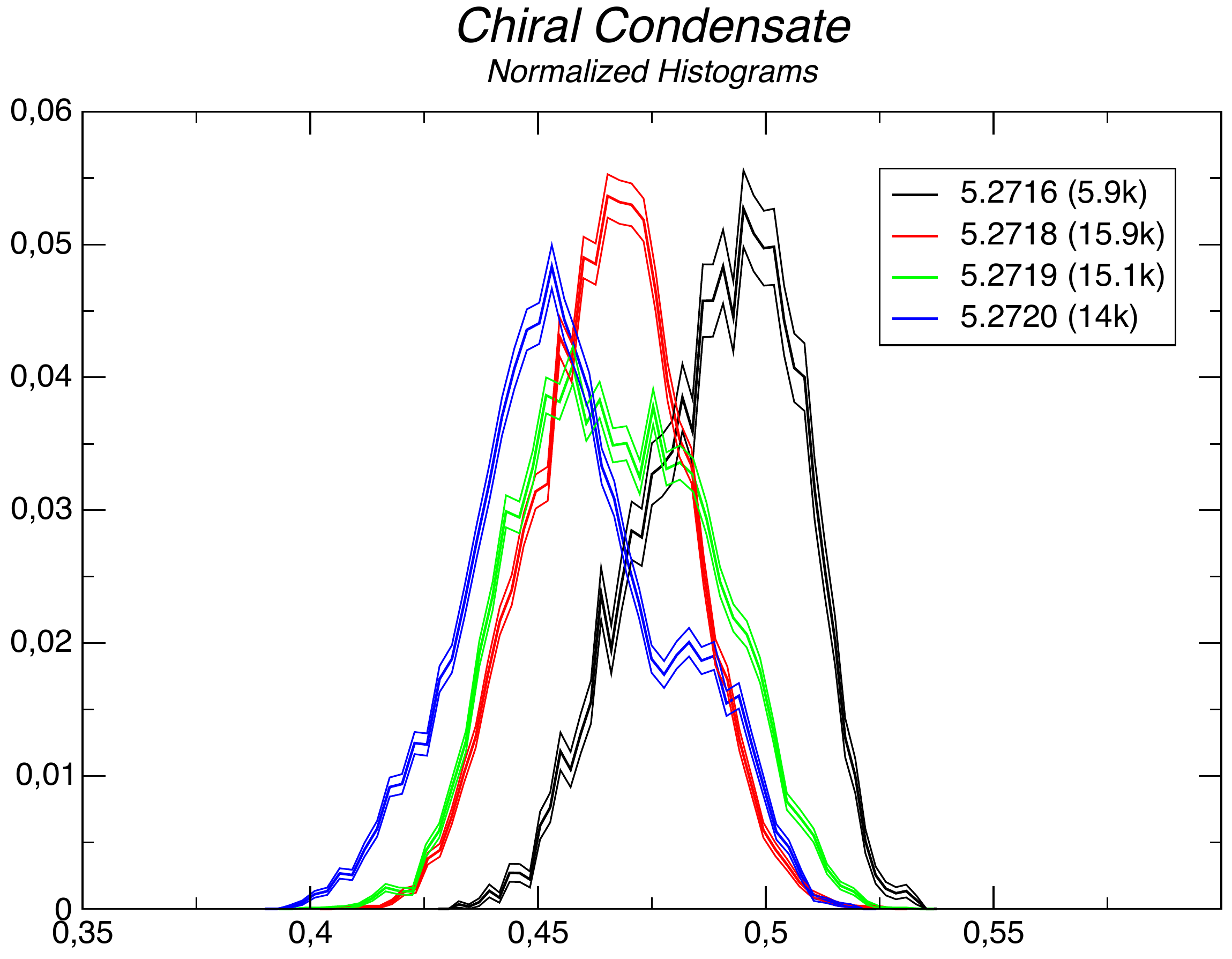}
\end{center}
\vspace{-6mm}
\caption{
{\bf (Left)} 
O(4) scaling test of the chiral condensate in two-flavor QCD with unimproved staggered quark action and unimproved plaquette gauge action on $N_t = 4$--12 lattices,
The curves are the O(4) scaling function \cite{Milc96}.  
{\bf (Right)} 
Histogram of chiral condensate in two-flavor QCD with unimproved staggered quark action and unimproved plaquette gauge action on an $N_t=4$ lattice \cite{Pisa}.
}
\label{Fig:O4Stag1}
\end{figure}

Besides the caveats, the results for the chiral transition with two flavors of staggered quarks have been quite confusing.
Although the early studies suggest that the transition is continuous in the chiral limit, the scaling behaviors observed there show neither O(2) nor O(4) \cite{Bielefeld94,Milc96,Jlqcd98}.
The left panel of Fig.~\ref{Fig:O4Stag1} is a comparison of QCD data with the O(4) [$\approx$ O(2)] scaling function by the MILC Collaboration \cite{Milc96}.
O(2) or O(4) scaling may appear only on very fine lattices and/or at smaller quark masses. 
(With a modified model including a four-fermi interaction, appearance of the O(2) scaling is reported at vanishing quark masses \cite{KogutPRD73}.)
In contradiction to these studies, the Genova-Pisa group reported indications of a first order transition on an $N_t=4$ lattice \cite{Pisa}.
A possible weakly two-state signal observed by them is shown in the right panel of Fig.~\ref{Fig:O4Stag1}.

\begin{figure}[tbh]
\begin{center}
  \includegraphics[width=0.66\textwidth]{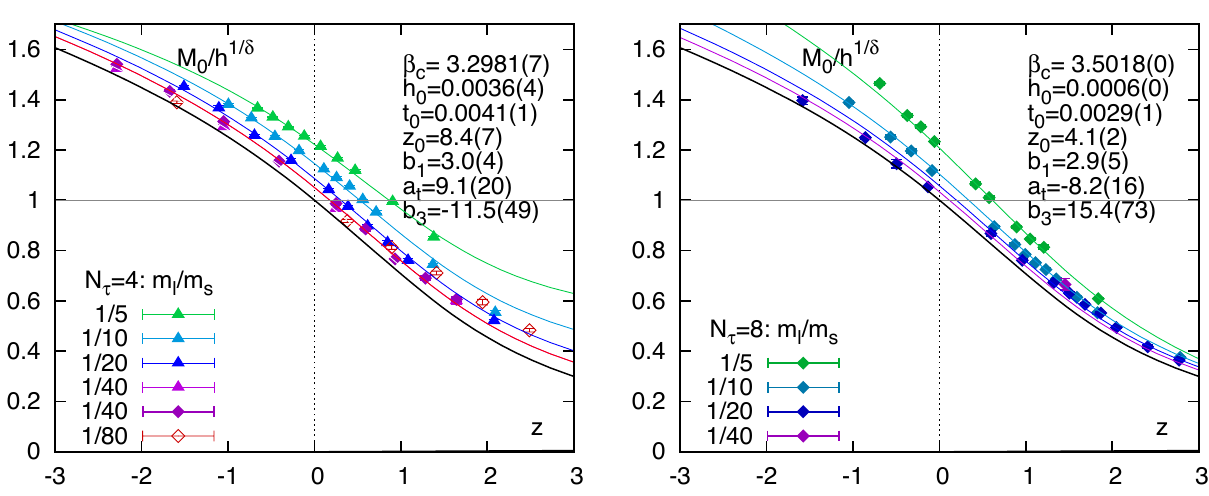}
\end{center}
\vspace{-6.5mm}
\caption{
Rescaled chiral condensate $M_0 = m_s \langle \bar\psi\psi \rangle_{ud}/ T^4$ 
in 2+1 flavor QCD with p4 staggered quark action and tree-level Symanzik gauge action 
on $N_t=4$ \cite{BNLB_O4} {\bf (Left)} and 8 \cite{SchmidtLat10} {\bf (Right)} lattices. 
The QCD data are fitted to the O(2) scaling function with an ansatz for deviations from the scaling.
}
\label{Fig:O4Stag2}
\end{figure}

All of these studies are made using the unimproved staggered quark action and the unimproved plaquette gauge action.
As discussed in Sec.~\ref{sec:Tc}, large taste-violation in these studies will make the quark masses effectively much heavier than those naively estimated by the pNG pion mass.

This year, a new study with improved staggered quark action was presented:
Adopting the p4 quark action and the tree-level improved Symanzik gauge action, 
the BNL-Bielefeld Collaboration performed O(2) and O(4) scaling fits in 2+1 flavor QCD on an $N_t=4$ lattice 
for the bare quark mass ratio in the range $m_{ud}/m_s = 0.05$ down to 0.0125
 with $m_s$ around the physical point \cite{BNLB_O4}.
 The quark mass range corresponds to $m_\pi^{\rm pNG} \approx 150$ MeV down to 75 MeV.
At the conference, preliminary results obtained on an $N_t=8$ lattice were presented too \cite{SchmidtLat10}.
The absence of spatial volume dependence suggests that the simulation points are in the crossover region.
For the first time with staggered-type quarks, O(2) [or O(4)] scaling has been observed.
At the same time, they found substantial deviations from the O(2) scaling function for $m_{ud}/m_s > 0.05$.
In Fig.~\ref{Fig:O4Stag2}, light quark chiral condensate is fitted to the O(2) scaling function assuming O(2) critical exponents, together with correction terms modeling the deviations from the scaling \cite{BNLB_O4}.
Different curves represent deviated scaling function for each quark mass.
They have tested the O(4) scaling too.
However, discrimination of O(2) from O(4) was not possible numerically.

Together with the adoption of improved actions, the substantial deviations from the O(2) scaling for $m_{ud}/m_s > 0.05$ explain the failures in previous studies.
A big impact of this study is the O(2) dominance around the physical point.
First, this suggests that the chiral transition in two-flavor QCD is second order.
Second, the $s$ quark mass for the tricritical point may be smaller than the physical value.
In Fig.~\ref{Fig:Nf21PhaseDiagram}, I have therefore put the physical point slightly higher than the previous plots.
The second observation is consistent with the fact that the region of first order transition around the three-flavor chiral limit has shrunk largely through the introduction of improved actions and the adoption of larger $N_t$ in the studies with staggered-type quarks.
When the parameter-dependence around the physical point is confirmed to be dominated by the O(4) exponents and the O(4) scaling function in the continuum limit, the studies of QGP will be largely accelerated.

\subsection{From the heavy quark limit}
\label{sec:HeavyQ}

\begin{figure}[tbh]
\begin{center}
  \includegraphics[width=0.355\textwidth]{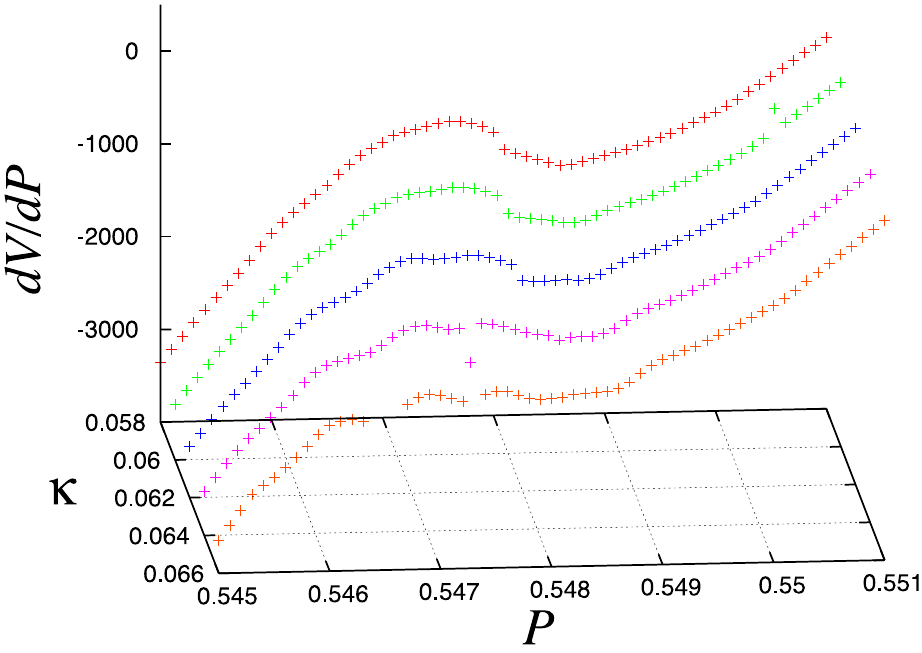}
  \includegraphics[width=0.27\textwidth]{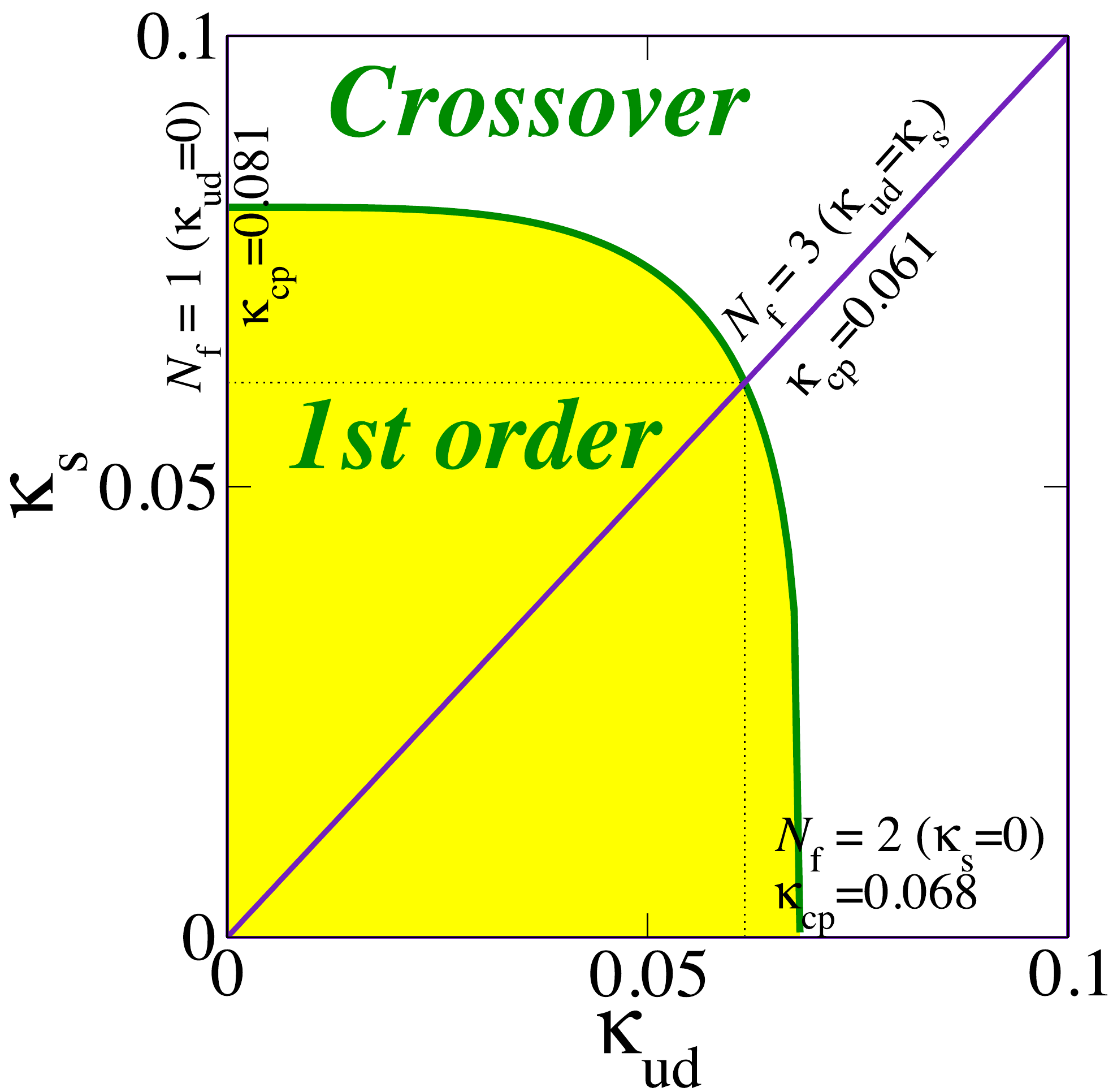}
 \hspace{0.5mm}
  \includegraphics[width=0.35\textwidth]{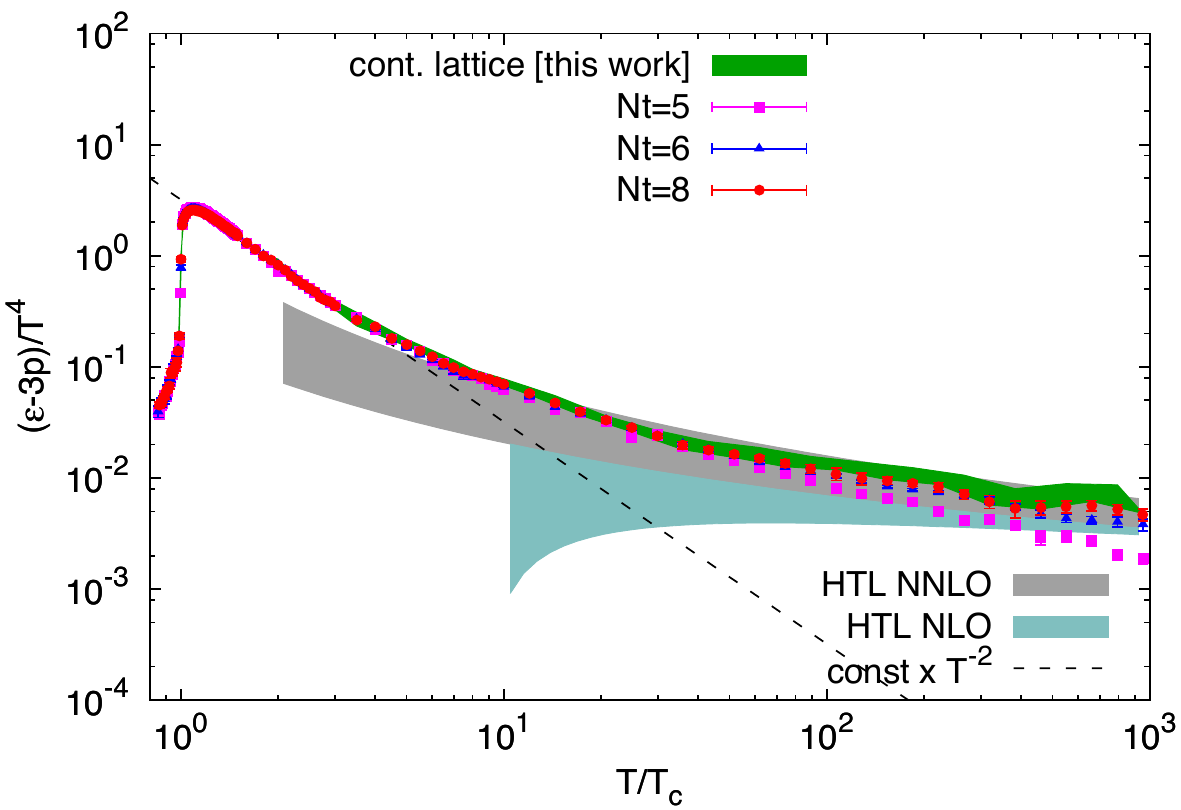}
\end{center}
\vspace{-5mm}
\caption{
{\bf (Left)}
Slope of an effective potential for the plaquette as a function of the hopping paremeter $\kappa$, obtained in two-flavor QCD with the standard Wilson quark action and the plaquette gauge action on an $N_t=4$ lattice \cite{SaitoLat10}.
{\bf (Center)}
The phase diagram for 2+1 flavor QCD around the heavy quark limit \cite{SaitoLat10}.
{\bf (Right)}
Trace anomaly in quenched QCD \cite{BorsanyiLat10}.
}
\label{Fig:HeavyQ}
\end{figure}

Before closing this section, let us turn our attention to the upper right corner of Fig.~\ref{Fig:Nf21PhaseDiagram}.
In the heavy quark limit, the deconfinement transition of SU(3) pure gauge theory is described by the first order transition of effective Z(3) Potts model \cite{YaffePRD26}. 
The phase of Polyakov loop serves the role of Z(3) spin.
Dynamical heavy quarks affect as an external magnetic field to the effective spins and break the Z(3) symmetry explicitly.
When we decrease the quark mass, the external magnetic field becomes stronger and eventually wipes away the first order transition into a crossover.

In the heavy quark limit, the Z(3) symmetry is spontaneously broken in the high temperature phase.
The deconfinement transition may thus be described by the percolation of Z(3) clusters.
Extending a previous study in the pure gauge theory \cite{GattlingerPLB690},
Danzer et al.\ studied the effect of dynamical quarks on the Z(3) clusters defined by the phase of local Polyakov loops \cite{DanzerLat10}.
The percolating nature at high temperatures turned out to be similar to the case of the pure gauge theory. 
However, the temperature dependence is much smoother than the pure gauge theory, in accordance with the crossover expected at the physical point.
Langelage, Lottini, Miura, Nakano and Ohnishi discussed about Polyakov loop effective theories in the context of a strong coupling expansion \cite{LangelageLat10, MiuraLat10}.

H.\ Saito reported a study on the fate of the deconfinement transition when the quark mass is decreased from infinity, adopting a method using probability distribution of observables combined with a reweighting technique \cite{SaitoLat10}.
They succeeded in calculating the effective potential, defined as the logarithm of the probability distribution for the plaquette, for a wide range of the plaquette expectation value in the heavy quark mass limit.
They then incorporated the effects of dynamical quarks by reweighting combined with the hopping parameter expansion.
The left panel of Fig.~\ref{Fig:HeavyQ} is their result for the slope of the effective potential as a function of the hopping parameter $\kappa$.
We see that the "S" shape typical for a first order transition gets weaker when $\kappa$ is increased ($m_q$ is decreased).
Determining the location of the critical point terminating the first order transition, they obtained the phase diagram shown in the center panel of Fig.~\ref{Fig:HeavyQ}.
As naively expected, the endpoint monotonically shifts to lighter quark mass as we increase the number of flavors.
The location of the end point is consistent with a previous estimation using an effective Z(3) spin model \cite{AlexandrouPRD60}.

Bors{\'a}nyi studied the EOS in quenched QCD up to extremely high temperatures to examine the onset of perturbative behavior \cite{BorsanyiLat10}.
Adopting an iterative technique and various improvements, they calculated the EOS up to $10000 \,T_c$ 
as shown in the right panel of Fig.~\ref{Fig:HeavyQ}.
They found that their result is consistent with a NNLO hard thermal loop perturbation theory above $8 \,T_c$.

\section{Equation of state}
\label{sec:EOS}

Calculation of the EOS in full QCD is demanding.
Besides the basic information such as the lattice scale, values of the beta functions, and  precise shape and location of the line of constant physics (LCP), we need to know the expectation values at $T=0$ to renormalize finite temperature observables at each simulation point.
Conventionally, we vary the coupling parameters along a LCP on a lattice with a fixed temporal lattice size $N_t$ to vary the temperature (fixed $N_t$ approach).
To calculate the EOS non-perturbatively, the integration method is adopted for which  precise data at lower temperatures are needed \cite{Integral}.
These require a systematic scan in a wide range of parameters on both zero and finite temperature lattices.
Furthermore, for a continuum extrapolation, we have to repeat the calculation at several values of $N_t$. 

Due to the big demand on the computer power, 
calculation of the EOS in realistic 2+1 flavor QCD has been made almost exclusively with staggered-type quarks.
This year, high precision results of the EOS for (almost) physical quark masses, identified using the pNG pion mass, were presented adopting different improved staggered quark actions,
though lattices are slightly coarser than those discussed in Sec.~\ref{sec:Tc}.
The first calculation of the EOS in 2+1 flavor QCD with Wilson-type quarks has been reported too, 
in which a new method was developed to beat the higher computational demand with Wilson-type quarks.

\subsection{EOS with 2+1 flavors of improved staggered quarks}

\begin{figure}[tbh]
\begin{center}
  \includegraphics[width=0.265\textwidth]{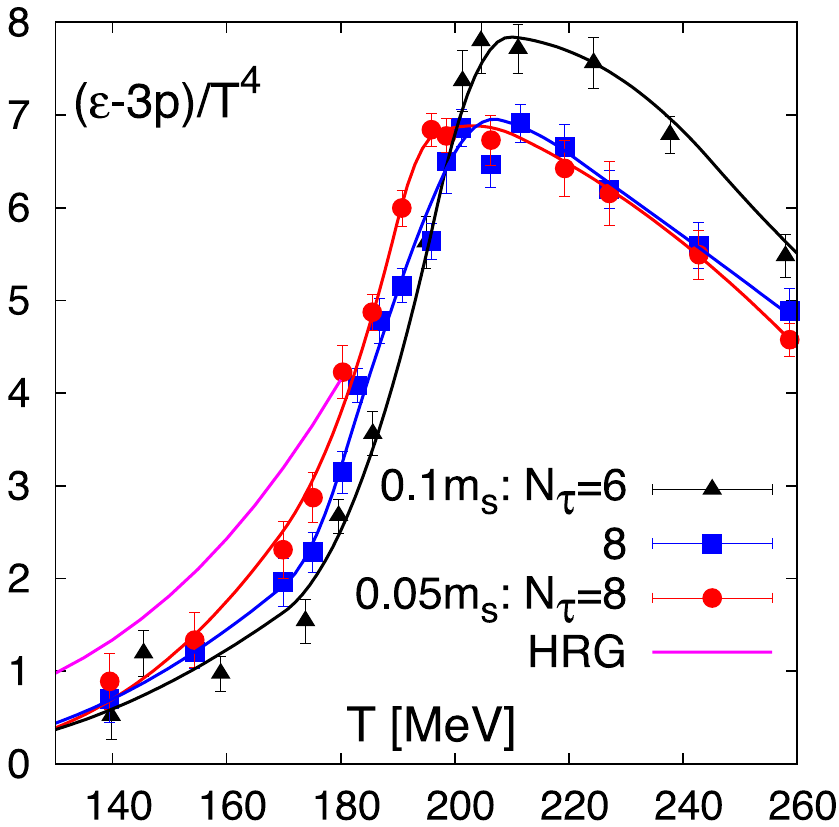}
 \hspace{4mm}
  \includegraphics[width=0.265\textwidth]{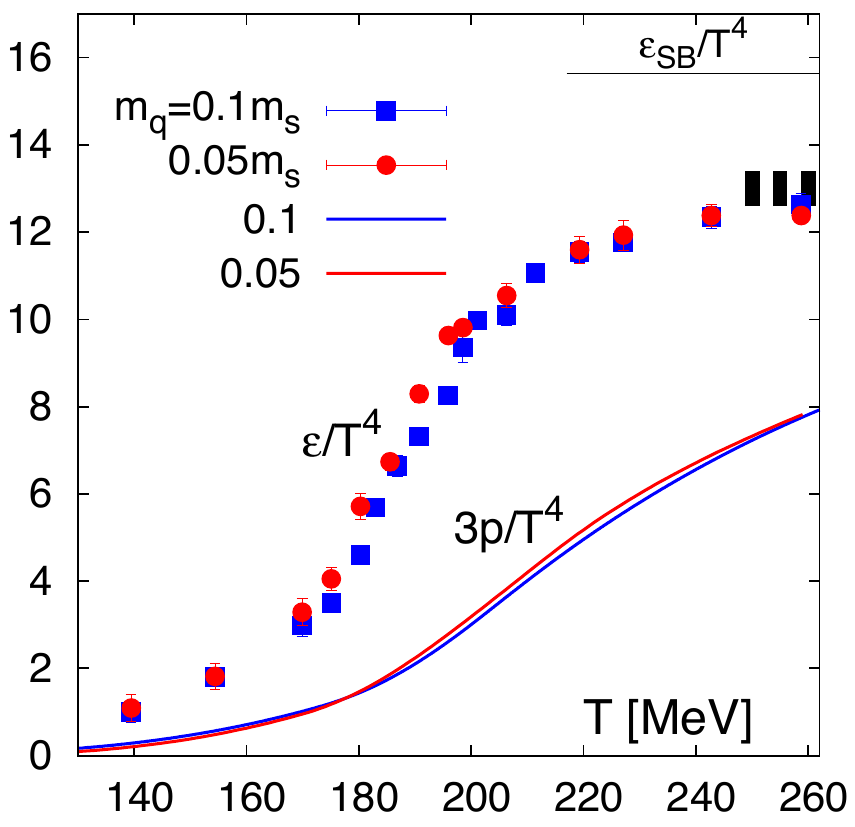}
 \hspace{4mm}
  \includegraphics[width=0.305\textwidth]{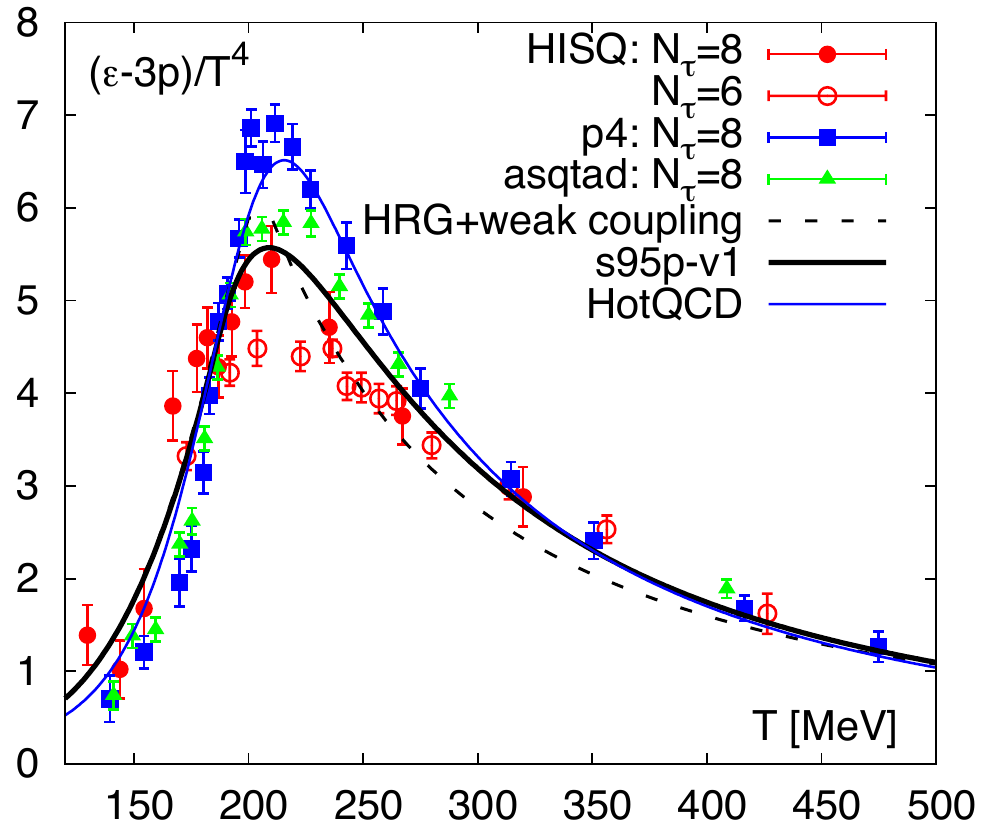}
 \end{center}
\vspace{-5.5mm}
\caption{
{\bf (Left, Center)} 
EOS in 2+1 flavor QCD at $m_\pi^{\rm pNG} \approx 154$ MeV ($m_{ud}/m_s = 0.05$) obtained on an $N_t=8$ lattice with p4 staggered quarks \cite{BNLB_EOSp}.
Results at $m_\pi^{\rm pNG} = 220$ MeV ($m_{ud}/m_s = 0.1$) are compared.
{\bf (Right)}
Comparison of the trace anomaly in 2+1 flavor QCD with HSQ, asqtad and p4 quarks at $m_{ud}/m_s = 0.05$ on $N_t=8$ lattices \cite{BazavovSoeldnerLat10}.
}
\label{Fig:EOSS1}
\end{figure}

The BNL-Bielefeld Collaboration published the EOS ``with physical quark masses'', based on their simulations using  p4 quarks on an $N_t=8$ lattice at the bare quark mass ratio $m_{ud}/m_s = 0.05$ with the $s$ quark mass fixed near the physical point \cite{BNLB_EOSp}.
Using a scale set by the Sommer scale $r_0=0.469$ fm, their simulation points correspond to $m_\pi^{\rm pNG} =154$ MeV, $m_K^{\rm pNG} = 486$ MeV and $m_{\eta_{s\bar{s}}}^{\rm pNG} = 663$ MeV with a precision of 3\%.
This study is an extension of their previous papers on the EOS with ``almost physical quark masses'' ($m_{ud}/m_s = 0.1$ where $m_\pi^{\rm pNG} = 220$ MeV) \cite{BNLB_EOSa,BNLB_EOSt8}.
In the left panel of Fig.~\ref{Fig:EOSS1}, the trace anomaly around the transition temperature is shown.
The shift of about $-5$ MeV from $m_{ud}/m_s = 0.1$ mentioned in Sec.~\ref{sec:Tc} is visible.
Adopting the integration method, the EOS shown in the center panel is obtained.
Note that the physical point was identified by the pNG pion mass, and the chiral and continuum extrapolations are not done yet.

\begin{figure}[tbh]
\begin{center}
  \includegraphics[width=0.285\textwidth]{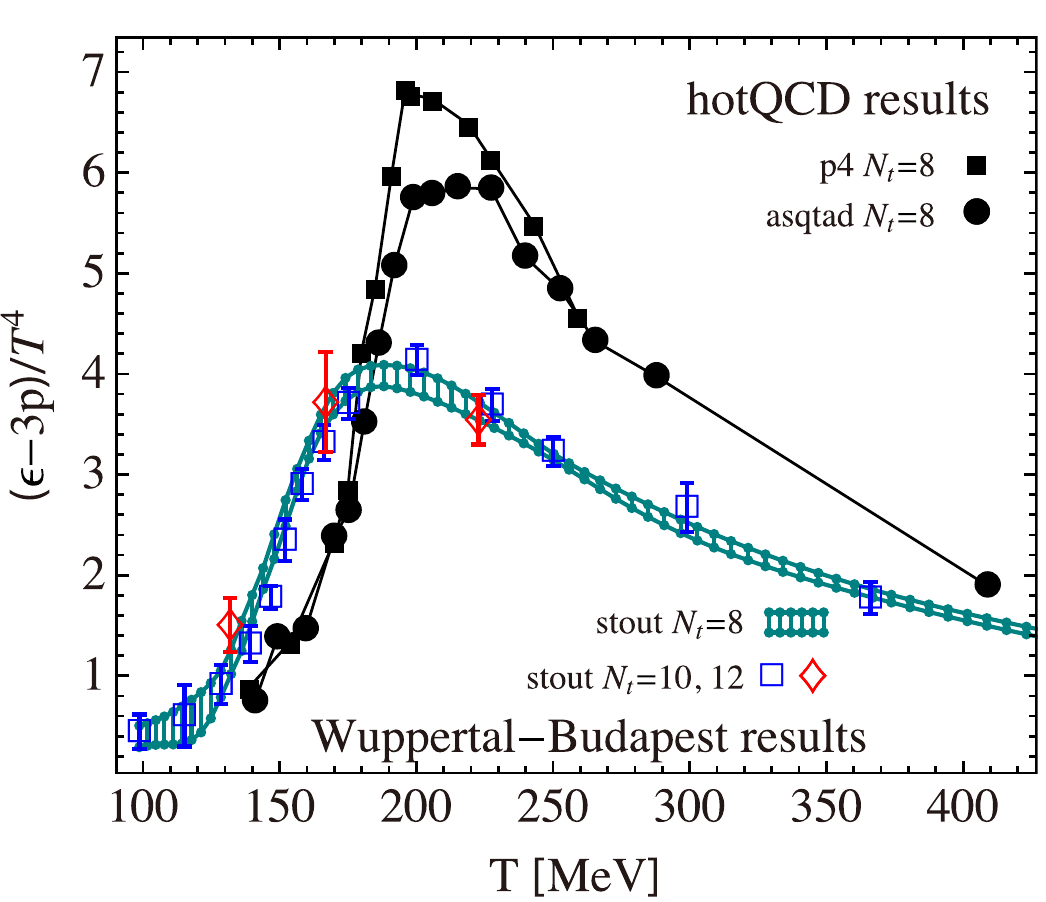}
\hspace{0.7mm}
  \includegraphics[width=0.335\textwidth]{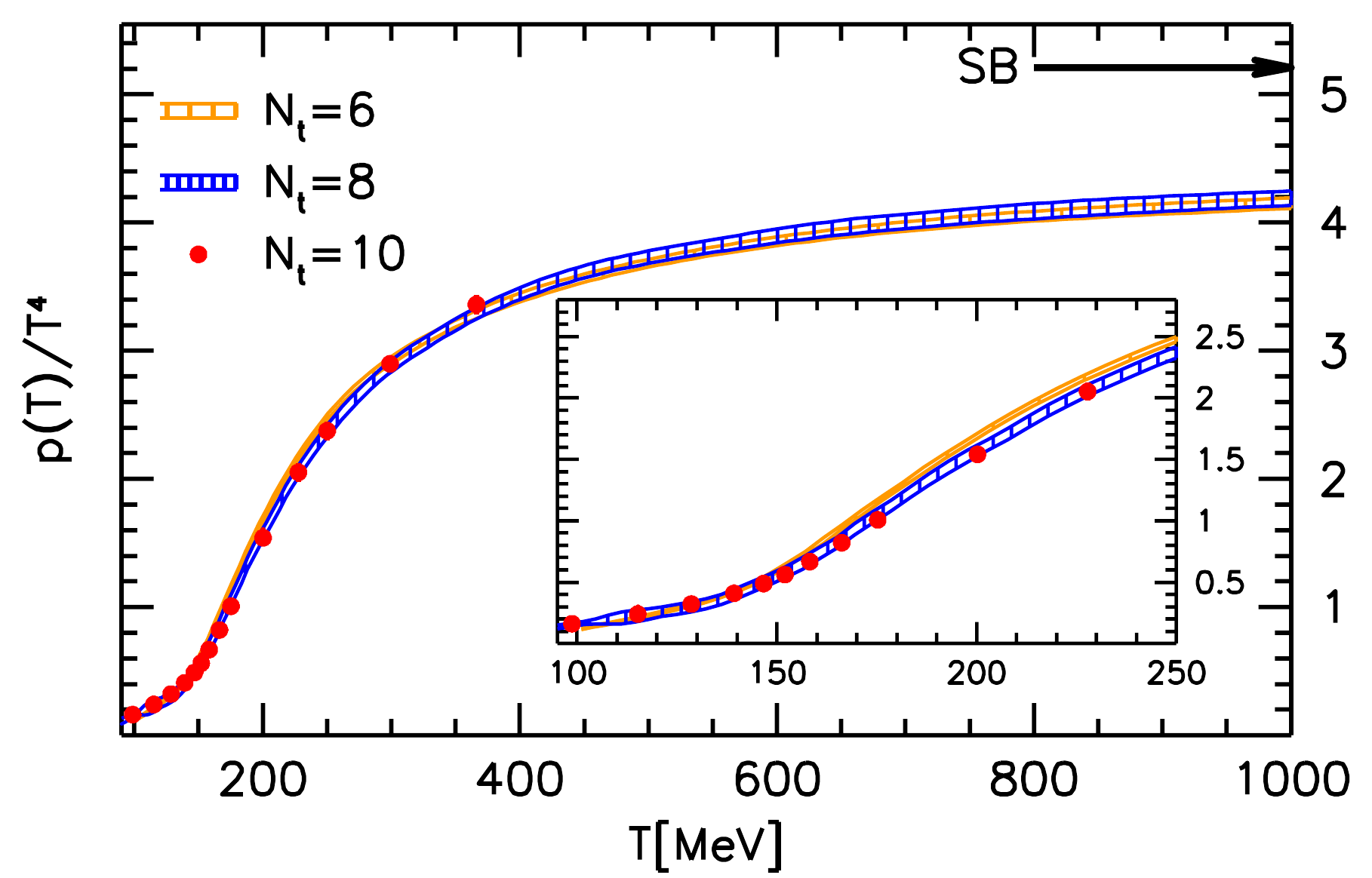}
\hspace{0.2mm}
  \includegraphics[width=0.335\textwidth]{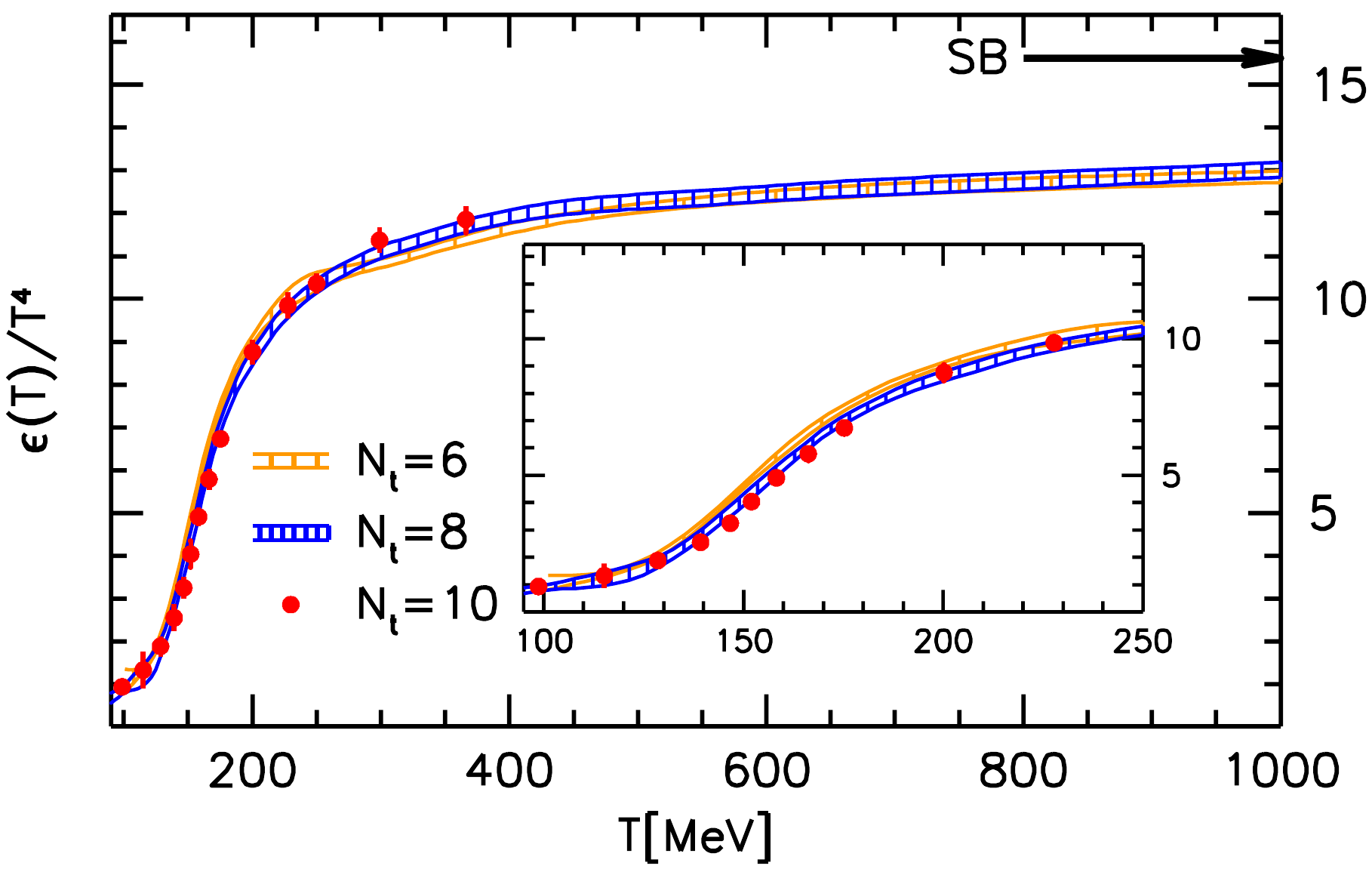}
 \end{center}
\vspace{-6.5mm}
\caption{
{\bf (Left)}
Trace anomaly in 2+1 flavor QCD with stout, asqtad and p4 quarks on lattices $N_t \ge 8$ \cite{WB_EOS10}.
The stout values are corrected by a tree-level improvement factor.
{\bf (Center, Right)}
EOS in 2+1 flavor QCD with stout quarks \cite{WB_EOS10}.
}
\label{Fig:EOSS2}
\end{figure}

At the conference, Bazavov and S\"oldner presented preliminary results of the HotQCD Collaboration for the EOS obtained with HISQ and asqtad quarks at $m_{ud}/m_s = 0.05$ on $N_t=8$ lattices \cite{BazavovSoeldnerLat10}.
In the right panel of Fig.~\ref{Fig:EOSS1}, the results of the trace anomaly are compared at  $m_{ud}/m_s = 0.05$ for HSQ, asqtad and p4 quarks.
The location of the transition is well consistent with each other, as discussed in Sec.~\ref{sec:Tc}.
On the other hand, the values of pressure and energy density are still moving.
We note that the peak of the trace anomaly around $T \sim 200$ MeV becomes lower with the adoption of a more improved action.

The Wuppertal-Budapest Collaboration presented the results for the EOS with stout quarks at the physical point, obtained on $N_t=6$--12 lattices \cite{WB_EOS10}.
In the left panel of Fig.~\ref{Fig:EOSS2}, trace anomaly is compared with those of p4 and asqtad quarks at $m_{ud}/m_s = 0.05$  on $N_t=8$ lattices.
The  stout values are corrected by a tree-level improvement factor. 
We note that the stout data indicates slightly smaller $T_c$ and much lower peak height of about 4 for $(\epsilon-3p)/T^4$.
A generalized integral method was adopted to calculate the pressure in a multi-parameter system.
They also attempted to calculate the EOS up to $T =1000$ MeV.
LCP up to large $\beta$ corresponding to this temperature was determined by a step-scaling method.
Their results for the EOS are shown in the center and right panels of Fig.~\ref{Fig:EOSS2}.

\subsection{EOS with 2+1 flavors of improved Wilson quarks}

About ten years ago, the CP-PACS Collaboration performed a systematic calculation of the EOS in two-flavor QCD with clover-improved Wilson quarks coupled to RG-improed Iwasaki glue \cite{CppacsPRD63,CppacsPRD64}.
Extension of the study to 2+1 flavor QCD was not straightforward due to the heavy computational demand for Wilson-type quarks.

To overcome the problem, Umeda et al.\ (WHOT-QCD Collaboration) developed a new approach: a fixed scale approach combined with the $T$-integration method \cite{Tintegral}.
In the conventional fixed-$N_t$ approach, the temperature is varied by varying coupling parameters on a LCP. 
In the fixed scale approach, on the other hand, the temperature is varied through a variation of $N_t$ with the coupling parameters fixed.
The conventional integration method \cite{Integral} in which the pressure is expressed as a line integral of the trace anomaly in the coupling parameter space is inapplicable in the fixed-scale approach because the simulations are performed only at a point in the coupling parameter space.
Therefore, they developed an alternative integration method, the $T$-integration method \cite{Tintegral} in which the pressure is given as an integration in $T$:
${\displaystyle
\frac{p}{T^4} = \int^{T}_{T_0} dT \, \frac{\epsilon - 3p}{T^5}
}$.

The fixed-scale approach is complementary to the fixed $N_t$ approach.
Towards the high $T$ limit, the fixed-scale approach suffers from lattice artifacts due to small $N_t$, while the fixed $N_t$ approach is applicable for local quantities which are insensitive to the system volume.
In the low $T$ region, the fixed $N_t$ approach suffers from coarseness of the lattice, while the fixed-scale approach keeps the lattice spacing $a$.
With the fixed scale approach, a LCP is automatically followed.
Since a common zero-temperature results can be used to renormalize thermodynamic observables at all temperatures, the cost for zero-temperature simulations can be largely reduced.
We may even borrow zero-temperature configurations on the ILDG..
[See Ref.~\cite{GavaiLat10} for a different application of the fixed-scale approach.]

\begin{figure}[tbh]
\begin{center}
  \includegraphics[width=0.3\textwidth]{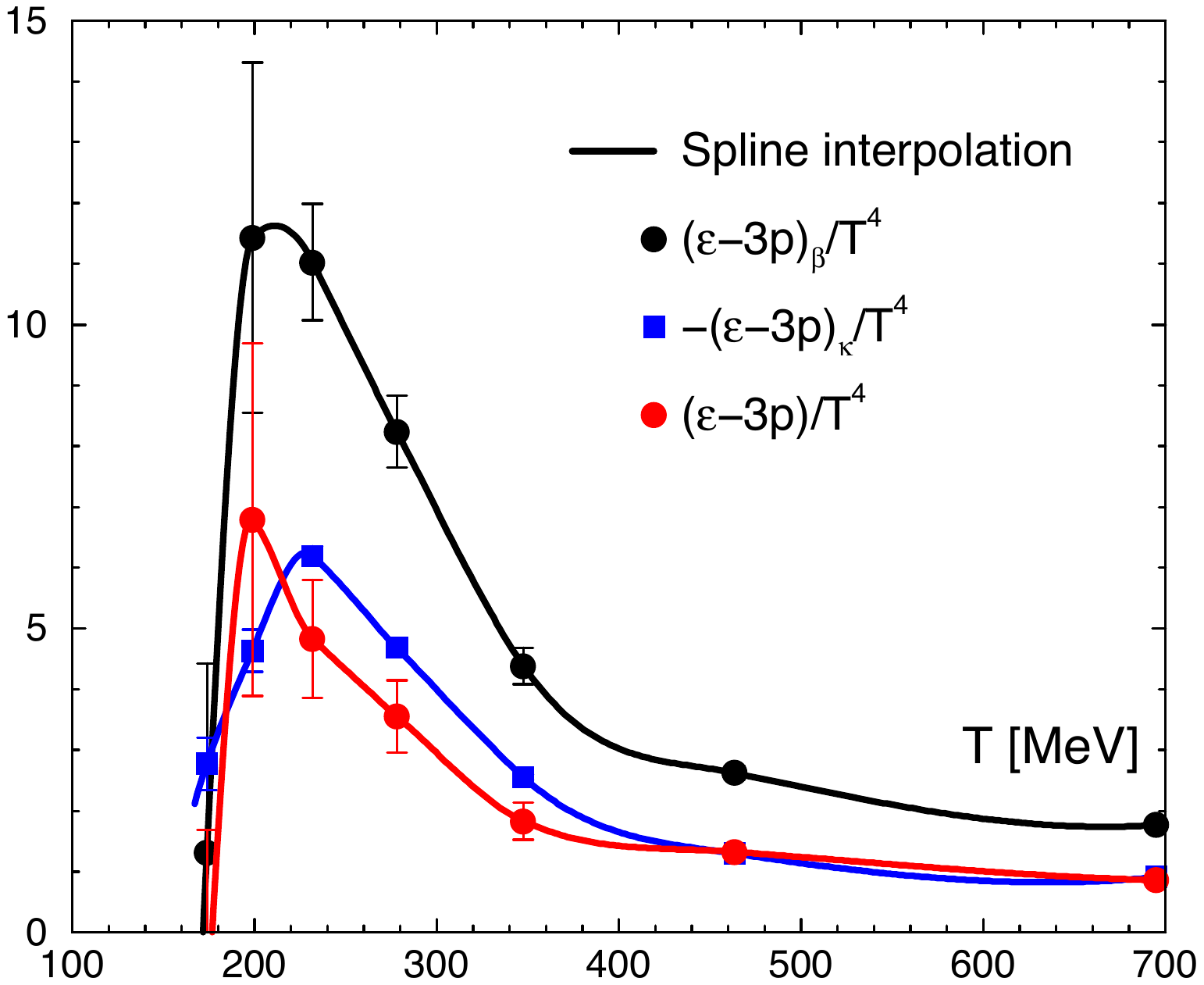}
  \hspace{10mm}
  \includegraphics[width=0.3\textwidth]{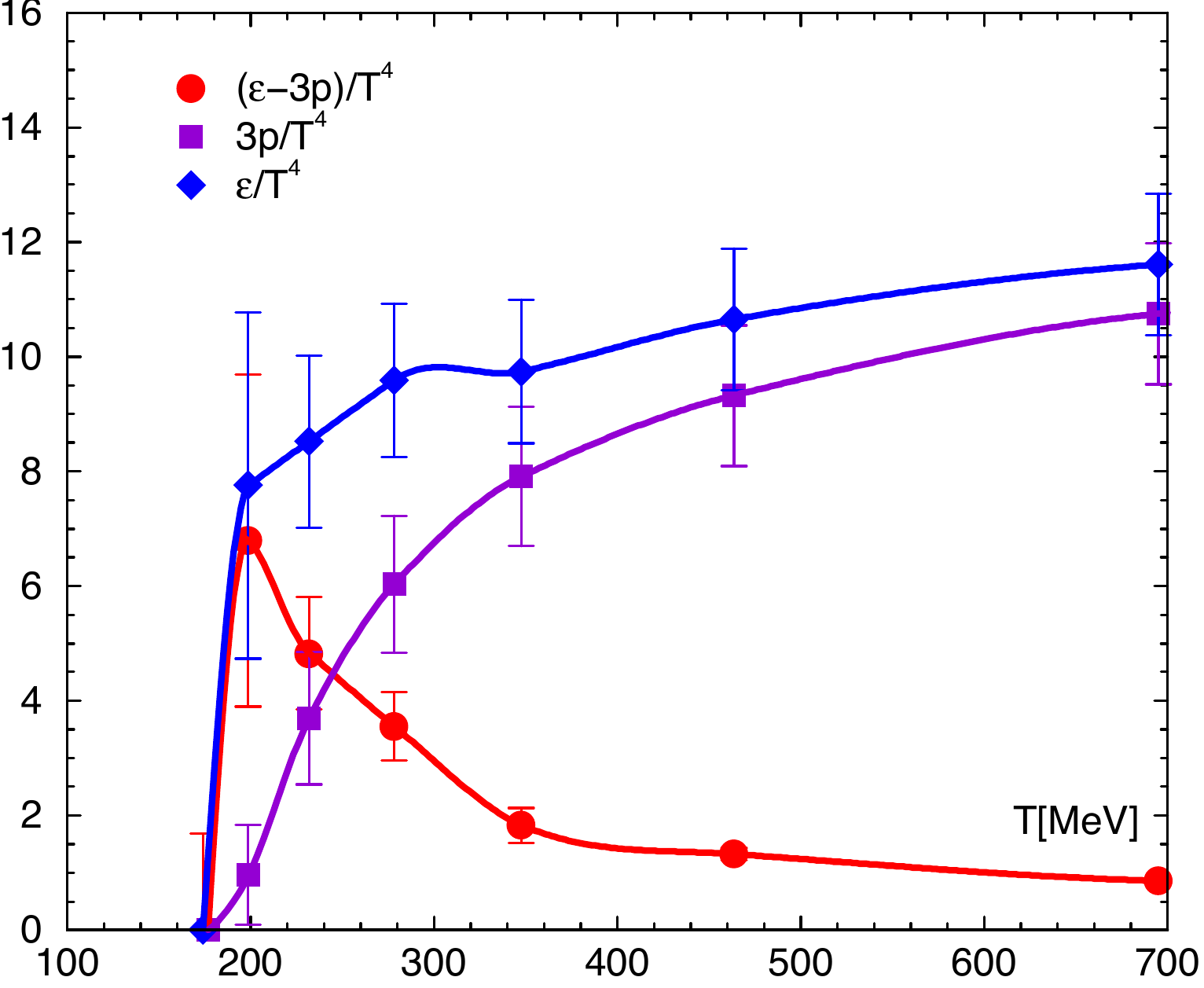}
\end{center}
\vspace{-5mm}
\caption{
EOS in 2+1 flavor QCD with clover-improved Wilson quarks adopting the fixed-scale approach \cite{UmedaLat10}.
{\bf (Left)}
Trace anomaly and its decomposition into contributions proportional to the beta function for $\beta$ and those for $\kappa$'s.
The latter contribution is multiplied by $-1$.
{\bf (Right)}
EOS obtained by the $T$-integration method.
}
\label{Fig:EOSW}
\end{figure}

Adopting the fixed-scale approach, Umeda et al.\ attempted the first calculation of the EOS in 2+1 flavor QCD with Wilson-type quarks \cite{UmedaLat10}.
They borrowed zero-temperature configurations generated by the CP-PACS+JLQCD Collaboration 
using the clover-improved Wilson quark action with non-perturbatively determined $c_{SW}$ and the RG-improved Iwasaki gauge action \cite{CppacsJlqcdPRD78}.
Choosing the finest lattice with the lightest $m_{ud}$, they carried out finite temperature simulations on $N_t=4$--16 lattices.
The left panel of Fig.~\ref{Fig:EOSW} is the result of the trace anomaly.
They found a big cancellation between the contributions proportional to the beta function for $\beta$  and for $\kappa_{ud}$ and $\kappa_s$.
The resulting low peak height is roughly consistent with those shown in Fig.~\ref{Fig:EOSS2} with highly improved staggered quarks, though the statistical error is large.
Note that, the peak in the left panel of Fig.~\ref{Fig:EOSW} is located at temperatures corresponding to $N_t \approx 14$.
From the $T$-integration using a trapezoidal interpolation of the trace anomaly, the EOS shown in the right panel of Fig.~\ref{Fig:EOSW} is obtained.
The dominant part of the errors are due to a large cancellation in the zero-temperature subtraction of the gauge action in the low temperature region.
Further statistics on low temperature lattices are needed  to reduce the error bars.

\section{Further hot issues}
\label{sec:HotIssues}

\subsection{Chiral magnetic effect}

\begin{figure}[tbh]
\vspace{1mm}
\begin{center}
  \includegraphics[width=0.345\textwidth]{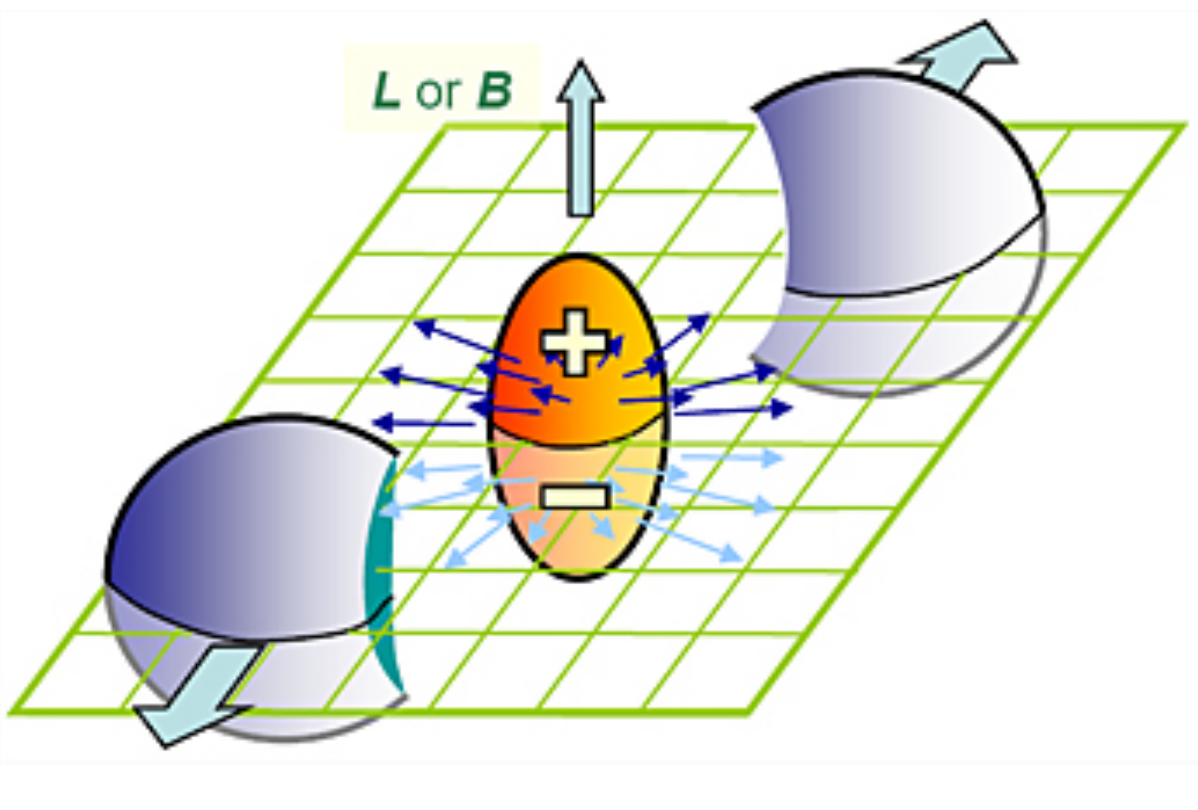}
  \hspace{0.1mm}
  \includegraphics[width=0.30\textwidth]{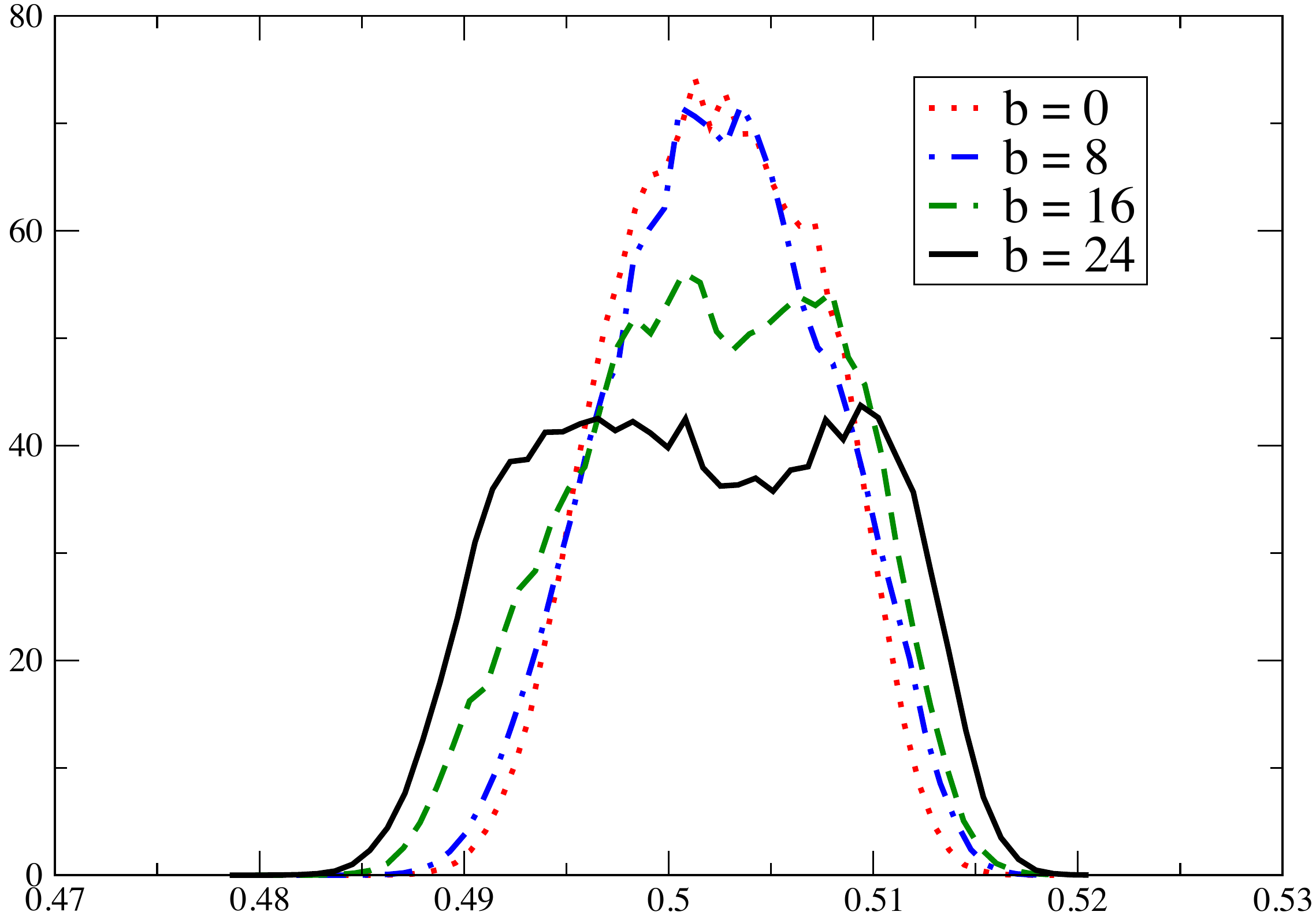}
  \hspace{0.5mm}
  \includegraphics[width=0.325\textwidth]{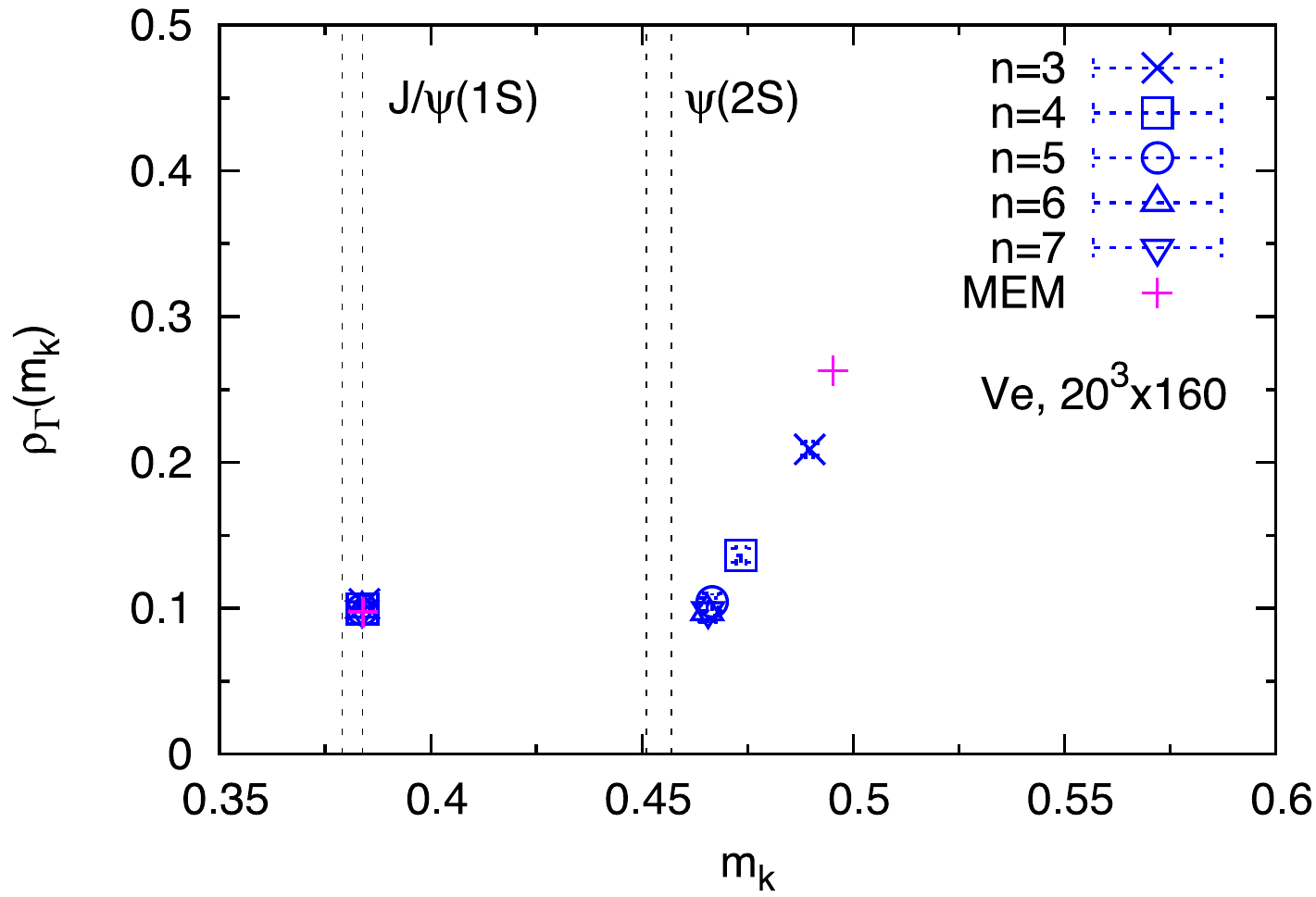}
\end{center}
\vspace{-5.5mm}
\caption{
{\bf (Left)} 
Strong magnetic field and charge asymmetry created at the collision point of heavy ion collisions \cite{STAR}
{\bf (Center)}
Plaquette histogram as a function of the external magnetic field $b$ in two-flavor QCD with staggered quarks on a $16^3\times4$ lattice \cite{DEliaLat10}
{\bf (Right)}
Charmonium spectral function for the vector channel at zero temperature obtained on a quenched anisotropic $20^3\times160$ lattice \cite{OhnoLat10}.
Blue symbols are the results of the variational method using different number of the basis functions.
The red ``$+$'' are the results of MEM.
Vertical dashed lines indicate the ranges of vector charmonia masses in nature.
}
\label{Fig:QGPmatter}
\end{figure}

Two nuclei colliding with a small offset in heavy-ion collisions can create quite strong magnetic field at the collision point \cite{RafelskiPRL36}. 
When the gluonic topological charge fluctuates locally, the strong magnetic field may yield an electric charge asymmetry with respect to the reaction place in the producted particles, as shown in the left panel of Fig.~\ref{Fig:QGPmatter}.
This apparent violation of parity is called the ``chiral magnetic effect'' \cite{CME}.
The STAR Collaboration reported an experimental evidence in favor of the predicted charge asymmetry last year \cite{STAR}.

Several lattice groups have studied the influence of external magnetic field on the quark matter.
Buividovich and Kalaydzhan presented a study of quenched SU(2) QCD with valence overlap quark under a strong background magnetic field \cite{BuividovicLat1}.
They observed an enhancement of electric current in the direction of the magnetic field, in accordance with the charge asymmetry.
They also found that a strong external magnetic field induces nonzero electric conductivity in the direction of the external magnetic field in the low temperature phase.
At the conference, D'Elia argued that, under a strong magnetic field, the transition temperature $T_c$ increases and the transition becomes sharper, as shown in the center panel of Fig.~\ref{Fig:QGPmatter}, based on a simulation of two-flavor QCD with unimproved staggered quarks at $N_t=4$ \cite{DEliaLat10}.

\subsection{Charmonium spectral functions}

Nonaka studied charmonium spectral functions using the maximal entropy method (MEM) \cite{MEM} on quenched anisotropic lattices with large spatial volume with and without momentum, and discussed the stability of $\eta_c$ and $J/\psi$ above the transition temperature \cite{NonakaLat10}.
Allton calculated non-zero momentum spectral functions by MEM, using twisted boundary conditions on quenched isotropic lattices \cite{AlltonLat10}. 
With staggered valence quarks, even time slice and odd time slice spectral functions are evaluated separately and combined to obtain the physical spectral function.
From the  longitudinal vector spectral function, they extracted the diffusivity.

Dilepton production may be a key process to explore the formation of QGP.
To calculate the dilepton rate and electrical conductivity without resorting to MEM, 
Francis and Karsch studied the vector meson correlation functions on large and fine quenched lattices in the high temperature phase  using clover valence quarks \cite{KarschLat10}. 
They performed continuum extrapolation of them and calculated dilepton rate and electrical conductivity.
Ding also presented their study of charmonium spectral functions by MEM, using clover valence quarks on quenched isotropic lattice \cite{DingLat10}.
To calculate a difference of spectral functions in which large energy parts are canceled and the low energy part we want to know is enhanced, they applied an ``extended MEM'' which is applicable for not positive-definite spectral functions.
Their result suggests that $J/\Psi$ is melt and $\eta_c$ starts to dissolve at $1.46 T_c$.

Ohno proposed a new method to calculate meson spectral functions adopting a variational method \cite{OhnoLat10}.
On finite volume lattices, the spectral functions should consists of discrete spectra only.
Therefore, a precise meaning of the continuous spectral functions obtained by MEM is not clear.
The variational method provides us with a method to directly evaluate the height and the location of low lying discrete peaks.
In the right panel of Fig.~\ref{Fig:QGPmatter}, the heights and the locations of the spectral function by the variational method are compared with those obtained by the conventional MEM for the ground and first exited states in the vector channel on a quenched anisotropic lattice at $T=0$.
For MEM, the area of the peak is identified with the height of discrete spectra.
The results for the ground state by the variational method fully reproduce the location and the area of the first peak obtained by the conventional MEM.
For the first exited state, the variational method converges to a different point with increasing the number of the basis operators, which is closer to the experimental mass.
Their preliminary result at $T>0$ indicates no clear signs of dissociation for $J/\Psi$ and $\eta_c$ up to $1.4 T_c$.

\subsection{Nature of hot quark matter}

Hydrodynamic evolution of quark matter plays an important role in the phenomenological analyses of heavy ion collisions.
New attempts to calculate transport coefficients were presented by Kohno \cite{KohnoLat10} and Maezawa \cite{MaezawaLat10} at the conference.

Banajee studied scalar mesons around the transition temperature with two flavors of unimproved staggered quarks on $N_t=6$ lattices with various spatial volumes \cite{BanajeeLat10}.
They found that scalar meson, which is known to be unstable at $T=0$, looks stable at $T = 0.94T_c$.
Kitazawa studied spectral properties of quarks in the Landau gauge in the deconfined phase on a quenched lattice with large spatial volume \cite{KitazawaLat10}.
His result suggests a minimum at finite momentum in the plasmino dispersion relation.

Three years ago, Garc{\'i}a-Garc{\'i}a and Osborn argued that the chiral transition of QCD may be viewed as a metal-insulator transition driven by the Anderson localization, based on a study low-lying Dirac eigenstates in quenched and 2+1 flavor QCD with 1-loop Symanzik gage action and and asqtad quarks \cite{GarciaGarcia_PRD75}.
A succeeding study of two flavors of  unimproved staggered quarks by the Indo-French Collaboration concluded that the localization of low-lying eigenstates seems to be a finite volume artifact \cite{IndoFrench}.
At the conference, Kov{\'a}cs presented a new study of quenched SU(2) QCD and argued that the results of the Dirac spectrum are in favor of the Anderson localization picture \cite{KovacsLat10}.
More works are needed to clarify the status.

\subsection{QCD-like theories}

At the conference, Datta presented their study of the EOS in SU($N_c$) pure gauge theories for $N_c=3$--6 \cite{DattaLat10}.
They concluded that the speculated ``strongly coupled conformal properties'' (the diagonal dashed line in the plot) are not seen.
Similar result was obtained by Panero for $N_c=3$--8.
The $N_c$-dependence turned out to be mild, and, by adjusting the parameters, good agreement with a prediction of a holographic QCD model expected in the large $N_c$ limit was discusses \cite{PaneroPRL103}.
Feo presented a study of the EOS in 2+1 dimensions for $N_c=2$--6
and argued that low-lying glueballs do not explain the trace anomaly data in the low temperature phase \cite{FeoLat10}. 
For QCD-like theories with various $N_c$, many studies in different focusing have been put on the archive last year.
An incomplete list includes \cite{LargeNc1,LargeNc2,LargeNc3}.

\section{Conclusions}

There were quite a few important advances in finite temperature QCD on the lattice in the past year.
Big progress was made, in particular, with improved staggered quarks:
The fruitful conflict about the value of the QCD transition temperature was (almost) settled.
The O(2) chiral scaling was finally observed by reducing the light quark mass down to a quite low value.
The EOS and thermodynamic quantities are being measured in 2+1 flavor QCD around the physical point.

From the intensive efforts to clarify the origin of the discrepancy in the transition temperature, we have learned the importance of the effect of taste violation in the simulations with staggered-type quarks.
By correctly handling it --- e.g.\ by adopting a highly improved action to sufficiently suppress the taste violation and by consulting the RMS hadron masses to judge the effective value of quark masses and to estimate the magnitude of taste violation effects --- simulations with staggered -type quarks may achieve a quite high precision.

On the other hand, the worry about the fundamental formulation of the staggered-type quarks remains.
Therefore, it is indispensable to cross check the results 
with other theoretically sound lattice quarks.
We have seen that steady advances are being made with Wilson-type quarks.
Calculation of the EOS in 2+1 flavor QCD has started with improved Wilson quarks.
Furthermore, simulations with the domain-wall quark action are now entering a level of quantitative investigations.
Lattice chiral quarks will open many applications beyond QCD.
To compare with the results from the staggered-type quarks, however, further reduction of light quark masses towards the physical point is necessary.

\vspace{2mm}
I thank the organizers of Lattice 2010 for stimulating conference and their supports.
I am grateful to D.\ Banerjee, A.\ Bazavov, P.\ Bicudo, S.\ Bors{\'a}nyi, F.\ Bruckmann, M.\ Cheng, M.\ Chernodub, S.\ Datta, M.\ D'Elia, C.\ DeTar, S.\ Ejiri, A.\ Feo, Z.\ Fodor, C.\ Gattlinger, F.\ Karsch, T. Kovacs, A.\ Maas, H.\ Ohno, M.\ Panero, O.\ Philipsen, H.\ Saito, K.\ Szabo, G.\ Schierholz, W.\ S\"oldner and T.\ Umeda for sending me their new results and comments. 
I also thank the participants of Lattice 2010 for useful discussions.
This work is in part supported by Grants-in-Aid of the Japanese Ministry of Education, Culture, Sports, Science and Technology (No. 21340049).


\end{document}